\documentclass{aastex62}

\usepackage{graphicx}                    
\usepackage{amssymb,amsmath}                    
\usepackage{color}                       
\usepackage{bm}
\usepackage{appendix}

\newcommand{\rmd}{ {\ \mathrm d} }
\renewcommand{\vec}[1]{ {\bm #1} }
\newcommand{\mvec}[1]{ {\mathbf #1} }

\newcommand{\rev}[1]{{\bf\color{blue}{}}}

\begin{document}

\title{Probabilistic Inversions for Time--Distance Helioseismology}

\author{Jason Jackiewicz}
	\affiliation{New Mexico State University, Department of Astronomy, Las Cruces, NM 88003, USA}
	\email{jasonj@nmsu.edu}

\begin{abstract}
  Time--distance helioseismology is a set of powerful tools to study localized features below the Sun's surface. Inverse methods are needed to robustly interpret time--distance measurements, with many examples  in the literature. However, techniques that utilize a more statistical approach to inferences, and that are broadly used in the astronomical community, are less-commonly found in helioseismology. This article aims to introduce a potentially powerful inversion scheme based on Bayesian probability theory and Monte Carlo sampling that is suitable for local helioseismology.  We first describe the probabilistic method and how it is conceptually different from standard inversions used in local helioseismology. Several example calculations are carried out  to compare and contrast the setup of the problems and the results that are obtained. The examples focus on two important phenomena that are currently outstanding issues in helioseismology: meridional circulation and supergranulation. Numerical models are used to compute synthetic observations, providing the added benefit of knowing the solution against which the results can be tested. For demonstration purposes, the problems are formulated in two and three dimensions, using both ray- and Born-theoretical approaches.   The results seem to indicate that the probabilistic inversions not only find a better solution with much more realistic estimation of the uncertainties, but they also provide a broader view of the range of solutions possible for any given model, making the interpretation of the inversion  more quantitative in nature. The probabilistic inversions are also easy to set up for a broad range of problems, and they can take advantage of software that is publicly available. Unlike the progress being made in fundamental measurement schemes in local helioseismology that image the far side of the Sun, or have detected signatures of global Rossby waves, among many others,  inversions of those measurements have had significantly less success. Such statistical methods  may help overcome some of these barriers to move the field forward.

\end{abstract}

%



\section{Introduction}

Inversions play a critical role for the interpretation of helioseismic measurements. In  global helioseismology, inversions of the frequency spectrum of the Sun's low-degree modes  have been used to determine its interior structure \citep{jcd1996}, including internal differential rotation \citep{thompson2003,howe2009b}. In the local framework of seismology, inversions of wave-packet  travel times,  or of ring parameters, are employed for measuring sub-surface flows \citep{komm2007}, such as meridional circulation \citep{giles1997,zhao2013,jackiewicz2015,rajaguru2015}, supergranulation \citep{zhao2003,svanda2012}, and velocity  structures in the vicinity of  sunspots \citep{couvidat2006,gizon2009,moradi2010}.

Helioseismic inversions estimate sub-surface quantities. Two popular classes of techniques used for this estimation are Regularized Least Squares (RLS) and Optimally Localized Averages (OLA) \citep{gough1991,jcd1993,pijpers1994,schou1994,corbard1997,jensen1998,jackiewicz2008,svanda2011,jackiewicz2012,korda2019}. These methods rely on inversions of large matrices that may suffer from numerical  instabilities when the  matrices are ill-conditioned, which they often are. Furthermore, the cost or misfit function to be  minimized may be very irregular in the parameter space, and strong regularization or smoothing often needs to be applied. Due to various tuning strategies, it is recognized that computing these inversions is sometimes as much ``art'' as science \citep{basu2016}.


An alternative framework to interpret observational data relies on Bayesian theory and statistics. In its simplest form, Bayesian inference combines prior information on a model and its parameters with observational data to produce a posterior probability distribution function (PDF hereafter) of the model parameters. The PDF represents the complete  solution to the inverse problem, and  all of the information is formulated in terms of probabilities. For this to work, one must know the statistical properties of the noise in the data. For helioseismology, these properties are typically well understood \citep{gizon2004,fournier2014}.

The Bayesian computation of the PDF spans the whole model space. In the case that the PDF is  Gaussian, then the inverse problem can be straightforwardly solved using the  methods described above to give a reasonable ``most-probable model.'' However, if the nature of the data or prior information is  complex, such that the  PDF  is not very smooth or is  multimodal, then a most-probable model has little meaning. In this case, it is important to  characterize the full shape of the PDF as to provide  realistic uncertainties on the estimations. The problem becomes one of sampling, rather than optimization. 

This is where methods of Markov Chain Monte Carlo (MCMC hereafter) come in. Modern MCMC techniques are actively being developed that efficiently and effectively sample multi-modal, multi-dimensional distribution functions of the parameter space. They work by drawing random samples that are distributed according to the properties of the PDF. Coupling these samplers to Bayesian inferences to solve problems is what will be referred to in this  article as \textit{probabilistic inversions}.



Apart from Earth seismology, which has a very mature MCMC inversion literature \citep[see the references within, and the references to][]{sambridge2002}, global helioseismology and asteroseismology have employed probabilistic methods much more sparsely. The applications have  not primarily  been for standard inversions either, but  for statistical measurements of the properties of individual seismic mode parameters (frequencies, amplitudes, linewidths) \citep[\textit{e.g.} the Diamonds package of][]{corsaro2014}.  Local helioseismology has seen even  less adoption. A notable exception is the current solar coronal seismology work led by I.~Arregui \citep[see][and references therein]{arregui2018}.  In other areas of astronomy, probabilistic inversions have proven to be a  robust way to interpret astronomical observations \citep{sharma2017}. Indeed, in a relatively recent article presenting a new MCMC Bayesian tool for the Python programming language, \citet{foreman2013} discuss its usage for general astronomical problems.  That publication has over 3000 citations in ADS (as of January 2020). A couple dozen are related to asteroseismology, but none to local helioseismology.

Therefore, we feel it could be useful to provide some examples of probabilistic inversions for local helioseismology. This article is written for people working in the field of solar physics and helioseismology who might not be very familiar with the utility of such techniques. We caution that there will be few details  in the derivation of Bayesian statistics and MCMC, so that more focus can be appled to  example tools and methods that can be used to solve certain classes of helioseismic problems. 

The rest of the article is organized as follows: In Section~\ref{sec:background},  the basic formulations of standard linear inversions and Bayesian inferences are described, as well as  how they are connected in certain cases. Section~\ref{sec:examples} provides examples of both types of inversions for two relevant problems in local helioseismology: inferring the flows of meridional circulation and those of supergranulation. This section compares in detail the results and outputs from the inversions. The final sections present a discussion of when one inversion technique might be preferable to another, and we end with a summary of the work presented. The appendices provide more details of the inversions, as well as a brief description of the probabilistic inversions in a simplified example to understand their usefulness at a conceptual level.


\section{Deterministic and Bayesian Inferences}
\label{sec:background}
\subsection{Formulation of Standard Helioseismic Inversions}
\label{sec:typ}

The majority of local helioseismic inversions published over the last decade or so rely on variants of the Optimally Localized Averages (OLA) method that was developed for terrestrial seismology by \citet{backus1968}. The most widely used form of this class of \textit{linear}, deterministic inversions may be the Subtractive OLA \citep[SOLA: ][]{pijpers1994,jackiewicz2008,svanda2011,jackiewicz2012,greer2016}. However, some recent studies have begun to employ full-waveform techniques that are very promising, yet very computationally demanding. The methods are iterative in nature and do not assume linearity between the response of seismic waves and the perturbation. Hanasoge and collaborators are at the forefront of this effort \citep{hanasoge2011,hanasoge2014,bhattacharaya2016}, which also has a  mature history in  terrestrial seismology.

In any case,  SOLA inversions essentially provide a way to infer the perturbation  one spatial location at a time. Unlike RLS-type algorithms, which try to find a best fit to the data, SOLA  forms linear combinations of the data (while minimizing the errors) that spatially localize the inference. The solution can critically depend on the tuning of certain parameters. These are not model parameters, but parameters that control the type of solution one desires. There are tradeoffs in the solution, such as those between spatial resolution and noise amplification, which are tunable. There are also parameters that allow for regularizing of possible ill-conditioned, large matrices. The choices of these parameters can be somewhat subjective and non-rigorous.

Standard derivations of the SOLA method are common in the literature \citep[\textit{e.g.}][]{svanda2011,jackiewicz2012,korda2019}. Here, a slightly modified version is presented that will connect to the probabilistic equations in Section~\ref{sec:bayes}.  We follow  closely the notation of \citet{tarantola2005}. Where appropriate, the relationship to standard inversion terminology is given in parenthesis with italicized text.

Assume any model can be described by  $\vec{m}(\vec{r})$, where $\vec{r}$ denotes space. By model, we mean the quantity that inversions are seeking,  such as the flow structure of a supergranule or the sound-speed profile under sunspots.  Consider a generalized discrete dataset $\vec{d}$ that  is linearly related to the model  through an integral equation
 \begin{equation}
   \vec{d} = g(\vec{m}),
   \label{eq:gm}
 \end{equation}
where $g$ is some functional that describes the physics of the problem. If such an equation exists, it will be called a generative model.  For now, this relationship will be given as
 \begin{equation}
   \vec{d} = \mvec{G}\vec{m},
 \end{equation}
 where $\mvec{G}$ \rev{is a matrix made up of} vector functions (\textit{sensitivity kernels}).  The true, but as of yet unknown, model is related to some set of observed data through
\begin{equation}
  \vec{d}_{\rm obs} = \mvec{G}\vec{m}_{\rm true},
  \label{eq:dobs}
\end{equation}
which we consider error free for simplicity. We want to obtain a good estimate $\vec{m}_{\rm est}$ of $\vec{m}_{\rm true}$ at some location, and we therefore assume that the estimator model is linearly related to the observed data as
\begin{equation}
\vec{m}_{\rm est} = \vec{w}^{\rm T}\vec{d}_{\rm obs},
\end{equation}
where the $\vec{w}$ are constants (\textit{weights}).  Defining some resolution operator (\textit{averaging kernels}) as
\begin{equation}
  \mvec{R} = \vec{w}^{\rm T}\mvec{G},
  \label{eq:aveker}
\end{equation}
gives
\begin{equation}
  \vec{m}_{\rm est} = \mvec{R}\vec{m}_{\rm true}.
  \label{eq:mtrue}
\end{equation}
This equation implies that the estimation that will be found is a smoothed \rev{or weighted} version of the true model, since with finite data $\mvec{R}$ will never be a delta function. 

The constants [$\vec{w}$] are computed by  minimizing a cost function
\begin{equation}
  \min \left| \mvec{R} - \mvec{I}\right|^2,
  \label{eq:cost}
\end{equation}
where $\vec{I}$ represents a delta function, but in practice it is something more reasonable (\textit{Gaussian target function}). Minimization with respect to the weights gives 
\begin{equation}
  \vec{w} = \left(\mvec{G}\mvec{G}^{\rm T}\right)^{-1}\mvec{G}.
\end{equation}
This expression shows that a (usually) large matrix inversion is necessary to compute (\textit{kernel convolution  matrix}). \rev{In a standard local helioseismic inversion, the convolution matrix can be of order $10^5\times 10^5$ elements, although various Fourier methods can help reduce this size \citep{jackiewicz2012}. Additionally, in practice this matrix may contain other quantities such as the noise covariance and any regularization terms required for a smooth solution.} 

Finally, once the weights are obtained, the estimate is given by
\begin{equation}
  \vec{m}_{\rm est} = \mvec{G}^{\rm T}\left(\mvec{G}\mvec{G}^{\rm
      T}\right)^{-1}\vec{d}_{\rm obs},
  \label{mest}
\end{equation}
and
\begin{equation}
  \mvec{R} =  \mvec{G}^{\rm T}\left(\mvec{G}\mvec{G}^{\rm T}\right)^{-1}\mvec{G}.
\end{equation}
Notice that the observations are only involved in the last step: the calculation of $\vec{w}$ is not conditioned on the data at all.

It is interesting to point out that the model estimate is likely not the true model (again using a finite amount of data). So it is reasonable to postulate that the true model may have a form in which it is  related to the estimated model, plus some arbitrary, properly scaled  model $\vec{m_0}$ of similar smoothness
\begin{equation}
  \vec{m} = \vec{m}_{\rm est} + (\mvec{I}-\mvec{R})\,\vec{m}_0.
  \label{eq:gen}
\end{equation}
This expression serves as  a general solution to the inverse problem \rev{\citep{backus1968}. Since $\mvec{G}((\mvec{I}-\mvec{R})\vec{m}_0)=0$, the same operation on Equation~\ref{eq:gen} gives
  \begin{equation}
    \mvec{G}\vec{m} = \mvec{G}\vec{m}_{\rm est} = \mvec{G}\mvec{R}\vec{m}_{\rm true} = \mvec{G}\vec{m}_{\rm true} = \vec{d}_{\rm obs},
  \end{equation}
which was shown in Equation~\ref{eq:dobs}.}


\subsection{Background to Bayesian inferences}
\label{sec:bayes}


Full discussions of MCMC in the general context of Bayesian theory and astronomy applications can be found in many places \citep[\textit{e.g.}][]{sharma2017,hilbe2017}. A particularly useful pedagogical treatment is given by \citet{hogg2018}. Here a simple overview is provided to guide the later discussion and examples.


Imagine we have $N$ measurements of some observable comprising a data set \rev{$\vec{d}=\{d_i \,|\, i = 1,\ldots,N\}$}, and each measurement has an uncertainty $\sigma_i$, which are \rev{considered independent and } normally distributed \rev{for simplicity}. Now assume that  we possess a generative model that can, in principle, make predictions of the data through the operation $g(\vec{m})$, as in Equation~\ref{eq:gm}.  $\vec{m}$ is a model made up of $M$ parameters \rev{$\vec{m}=\{m_i \,|\, i = 1,\ldots,M\}$}. If many repeated measurements are made, then the expected frequency distribution (probability) of datum $d_j$ is
\begin{equation}
  p(d_j|\vec{m},\sigma_j) = \frac{1}{\sqrt{2\pi\sigma_j^2}}\exp\left[-\frac{(d_j - g_j(\vec{m}))^2}{2\sigma_j^2} \right].
\end{equation}
The vertical bar $|$ is read as ``given,'' so this expression is the probability of the datum $d_j$ \textit{given} the model and the uncertainty on $d_j$. Clearly, if the operation
\begin{equation}
  g_j(\vec{m}) \equiv \sum_{i=1}^M g_j(m_i)
\end{equation}
gives a number far from $d_j$, the resulting probability will be small. One wishes to maximize the probability, not just of one data point, but the entire set of observations. This is usually referred to as the  likelihood function \rev{[$L$]}, which is a product of individual probabilities
\begin{equation}
  L(\vec{d}|\vec{m},\vec{\sigma}) = \prod_{j=1}^N p(d_j|\vec{m},\sigma_j).
  \label{eq:like1}
\end{equation}
In practice, one may stop here and find the parameters that maximize the likelihood function, or, more conveniently, minimize the negative logarithm of it. The problem then reduces to least-squares fitting. The resulting model, identified from the probability of the \textit{data given the parameters}, is interpreted, however,  as the likelihood of the \textit{parameters given the data}.  This interpretation presents a formal inconsistency.

Bayes's Theorem can be easily derived from sum and product rules of probability theory. The result has four quantities.  One quantity is the likelihood function in Equation~\ref{eq:like1}. Another is any prior information [$I$] that we possess on the model and the uncertainties, which will be denoted $\rho(\vec{m},\sigma|I)$. The third is the evidence, [$p(\vec{d}|I)$], which is a effectively a normalization term and will not be important for our discussion. The final ingredient is the posterior probability distribution function, which is computed as
\begin{equation}
  {\rm PDF}(\vec{m}|\vec{d},\sigma, I) = \frac{L(\vec{d}|\vec{m},\vec{\sigma})\rho(\vec{m},\sigma|I)}{p(\vec{d}|I)},
  \label{eq:bayes}
\end{equation}
and defines Bayes's Theorem.  This important quantity is the statistical probability of the model \textit{given} the data, uncertainties, and any prior knowledge.

The model PDF is therefore related to the likelihood function, and it will closely resemble it if the priors are not very specific or informative. In this case the interpretation stated above is not so fatal.
However, some of the power of the Bayes framework is that if the prior knowledge of model parameters is non-trivial, then the PDF is too, and its complexity requires more sophisticated inference methods  to be applied. Priors  also restrict the parameter space to a smaller region than a likelihood function alone can.

In general, and in the examples below, the likelihood function is a multivariate normal distribution
\begin{equation}
  L(\vec{d}|\vec{m},\vec{\Sigma}) = \frac{1}{\sqrt{(2\pi)^k|\vec{\Sigma}|}}\exp\left[-\frac{1}{2}\left(\vec{d}-g(\vec{m})\right)^{\rm T}\vec{\Sigma}^{-1}\left(\vec{d}-g(\vec{m})\right)\right],
  \label{eq:like}
\end{equation}
where $\vec{\Sigma}$ is the data covariance matrix, $|\vec{\Sigma}|$ is its determinant, and $k$ is the dimension of the problem (length of $\vec{d}$).



To summarize, probabilistic inversions use Equation~\ref{eq:bayes} to compute the posterior PDF --  the joint probability distribution of parameters that is consistent with the data. The PDF will rarely have an analytic form, and it is not necessarily well-behaved or uni-modal. The goal is not to optimize (maximize) the PDF, or find its peak, but to know the whole distribution and sufficiently sample it. One can imagine one strategy, which would be looping over a (uniform) grid of parameter values and computing the resulting PDF. However, for high-dimensional problems, this would be extremely expensive and inefficient. Too many low-probability realizations would be calculated. Fortunately, there are alternative approaches.  There is a robust literature of different probability distribution sampling methods, but  most modern ones rely on MCMC techniques. The  basic difference among these methods is  how the sampler ``moves'' through the parameter space;  \textit{i.e.} how an algorithm decides to choose trial parameter values, so  that it hopefully spends more time in high-probability space. MCMC uses random numbers to drive the process. Metropolis--Hastings is one of the simplest and well-known algorithms \citep{press2007}.


A recently developed MCMC algorithm,  affine-invariant sampling \citep{goodman2010}, is what we adopt in this work.  This method is in a class of ensemble MCMC, since multiple chains, called ``walkers,'' are all in execution simultaneously as they explore the parameter space. The walkers can therefore be  run in parallel, but they are allowed to interact in certain ways to adapt the proposal densities and maintain their Markov properties.  It is a promising tool for sampling PDFs that are not extremely complex \citep{foreman2013}.


The connection of the SOLA inversion described in Section~\ref{sec:typ} to the probabilistic language where priors and covariances are considered is useful, and it can be made quite easily. Consider a linear least-squares problem. Including model priors, one  could construct a cost function (or least-squares function, or $\chi^2$-function) \rev{called $S$} as
\begin{eqnarray}
  2S(\vec{m}) &=&  (\vec{d}_{\rm obs} - \mvec{G}\vec{m} )^{\rm T}\mvec{C}_{\rm D}^{-1} (\vec{d}_{\rm obs}-\mvec{G}\vec{m})+\\
  &+& (\vec{m}-\vec{m}_{\rm prior})^{\rm T}\mvec{C}_{\rm M}^{-1}(\vec{m}-\vec{m}_{\rm prior}).
\end{eqnarray}
 $\vec{C}_{\rm D}$ and $\vec{C}_{\rm M}$ are the covariance matrices of the data and model priors (if known), respectively.  As outlined above, the  Gaussian posterior PDF can be computed from the cost function \rev{$S$} that has a form $\sim \exp(-S(\vec{m}))$. The center of the distribution, \textit{i.e.} the most likely model of the Gaussian PDF (the  model that minimizes the cost function) $\tilde{\vec{m}}$ and its covariance $\tilde{\vec{C}}_{\rm M}$ can be computed by  differentiation and shown to be
\begin{eqnarray}
  \tilde{\vec{m}} &=& \vec{m}_{\rm prior} + \mvec{C}_{\rm M} \mvec{G}^{\rm T}(\mvec{G}\mvec{C}_{\rm M}\mvec{G}^{\rm T} + \mvec{C}_{\rm D})^{-1}(\vec{d}_{\rm obs} - \mvec{G}\vec{m}_{\rm prior}), \\
  \tilde{\mvec{C}}_{\rm M} &=& \mvec{C}_{\rm M} -  \mvec{C}_{\rm M}\mvec{G}^{\rm T}(\mvec{G} \mvec{C}_{\rm M}\mvec{G}^{\rm T}+ \mvec{C}_{\rm D})^{-1}\mvec{G} \mvec{C}_{\rm M}.
\end{eqnarray}

In the SOLA method, there are no priors  in the model space. In the probabilistic language, this  implies  white noise with no correlations and a (possibly) infinite variance of the priors:
\begin{equation}
  \mvec{C}_{\rm M} \approx k\mvec{I}\;\;\quad (k\rightarrow\infty).
\end{equation}
Using the definitions in Section~\ref{sec:typ}, the posterior centers then reduce to
\begin{eqnarray}\nonumber
  \tilde{\vec{m}} &=&  \mvec{G}^t(\mvec{G}\mvec{G}^t)^{-1}\vec{d}_{\rm obs} +(\mvec{I}-\mvec{R})\vec{m}_{\rm prior},\\
  &=& \mvec{R}\,\vec{m}_{\rm true} + (\mvec{I}-\mvec{R})\vec{m}_{\rm prior}, \\
  \tilde{\mvec{C}}_{\rm M} &=& (\mvec{I}-\mvec{R})\mvec{C}_{\rm M}.
\end{eqnarray}
The first equation is precisely Equation~\ref{eq:gen}, a general solution of a SOLA inversion  with the prior information replacing the arbitrary model $\vec{m}_0$. In the SOLA language, if $\mvec{R}\approx\mvec{I}$ (a $\delta$-function like \textit{averaging kernel}), then $\tilde{\vec{m}} = \vec{m}_{\rm est}\simeq\vec{m}_{\rm true}$. In the probabilistic language, this implies there are no uncertainties in the posterior solution, so $\tilde{\mvec{C}}_{\rm M}\simeq \vec{0}$.

These two conclusions are identical, showing that, in principle, the  methods can arrive at similar results, yet only in ideal circumstances. What is hopefully demonstrated throughout the rest of this article is that the probabilistic method, in practice, is robust, practical, and gives  more \rev{realistic} uncertainties.






\section{Examples for Time--Distance Local Helioseismology}
\label{sec:examples}

The forward problem in  time--distance helioseismology is symbolically
formulated as \citep[e.g.][]{kosovichev1997,gizon2002,gizon2004}
\begin{equation}
  \delta\tau = \int_\odot K \delta q\, \rmd r,
  \label{eq:forward}
\end{equation}
where the travel-time shifts [$\delta\tau$] between two surface locations are caused by some (small) interior perturbation $\delta q$. The sensitivity  kernels [$K$]  mediate this
relationship, which is  considered to be linear. Any inversion consists of using the observed surface $\delta\tau$ and computed $K$ to find the unknown $\delta q$. In SOLA  methods, $\delta q$ is inferred at each spatial location, or at least one depth at a time. In probabilistic inversions, $\delta q$ must first be parametrized by some number of free parameters. The parameters are estimated using Bayes's theorem and MCMC, and then $\delta q$ can be studied over the whole domain.

We elucidate two rather simple examples of inversions based on common research areas in local
helioseismology. We compare the probabilistic inversions with the SOLA
method and contrast the computational particulars. We will only consider
examples of flows, and therefore travel-time differences are the important observables.



It is important to keep in mind that in what follows we are not solving any real problem. In one case, we are only inverting
synthetic observations that are computed in the forward sense from Equation~\ref{eq:forward}.  This does not tell us anything about the accuracy of the sensitivity kernels. They could be completely wrong. It only tells us about the inverse process, which is the goal here.
In the other case, inversions of a realistic numerical model are shown.  
Most helioseismic studies employ two ways of modeling the interaction
of seismic waves with inhomogeneities: ray theory or Born theory. Our
examples span these two cases.





\subsection{Meridional Circulation in a Ray-Theory Approach}

\subsubsection{The Toy Problem}

\begin{figure}
  \centerline{
    \includegraphics[width=.45\textwidth,clip=]{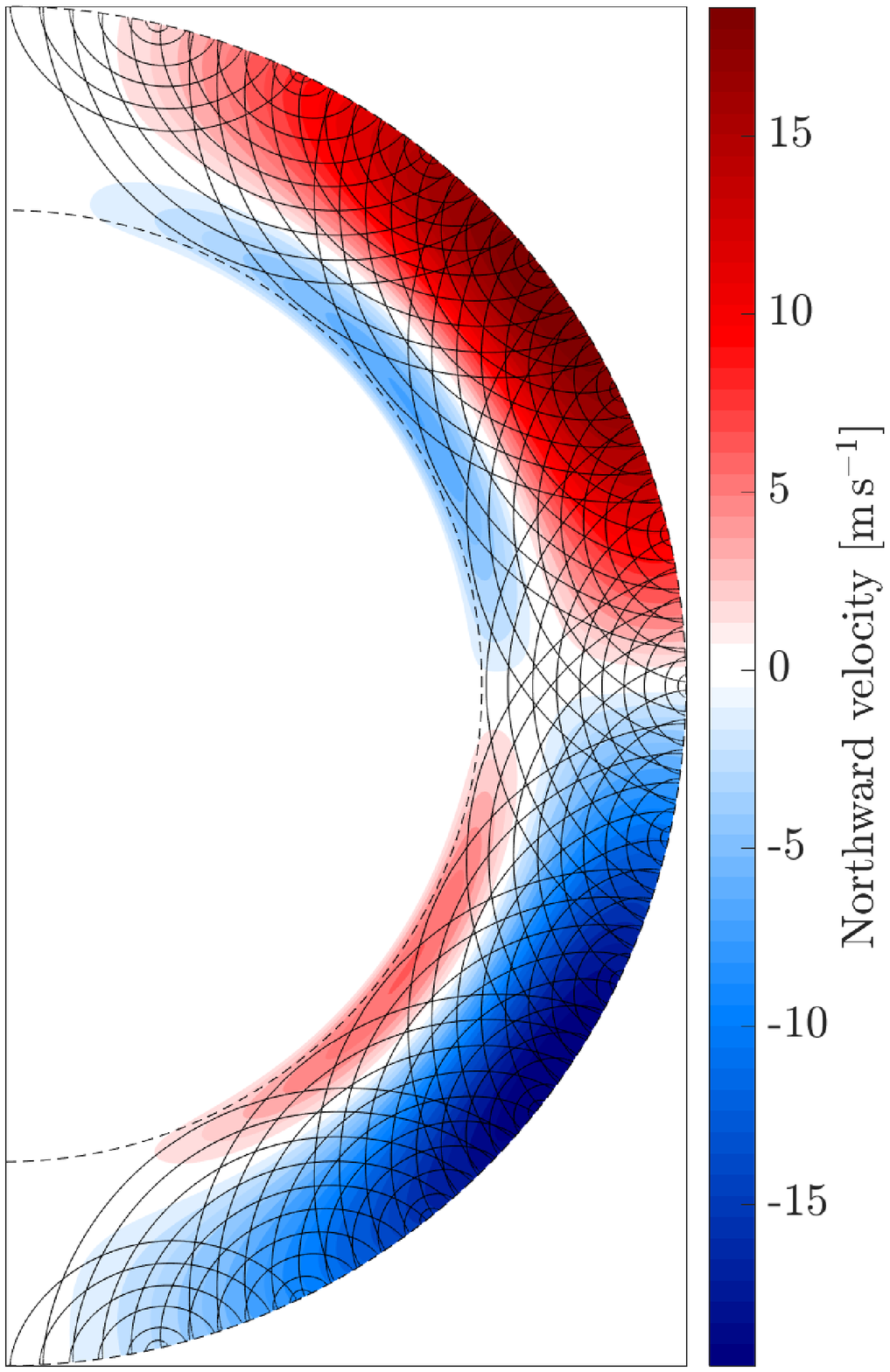}
    \includegraphics[width=.55\textwidth,clip=]{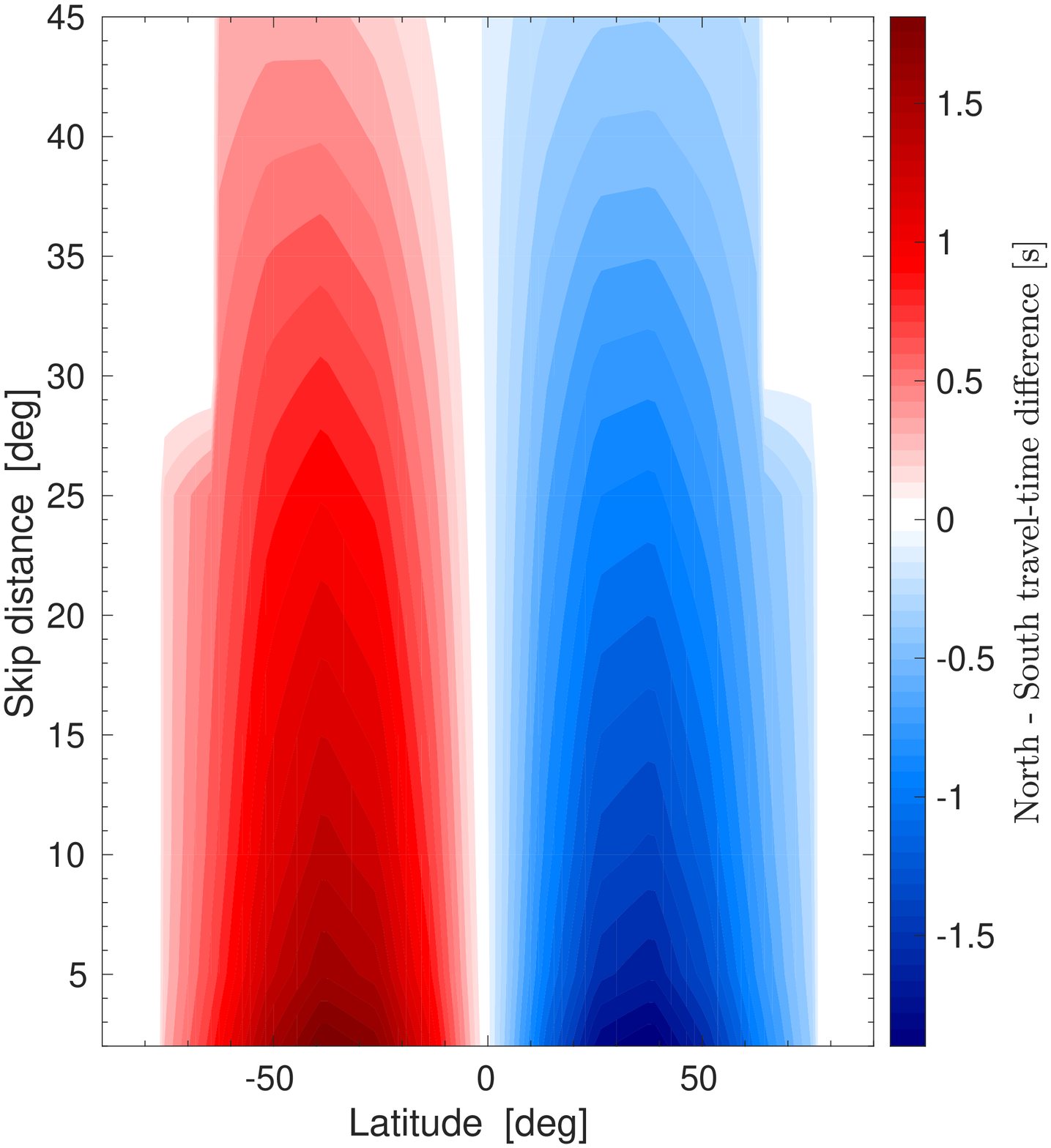}}
  \caption{Left: Input model latitudinal flow profile and ray paths. The color scale shows the northward velocity of a model computed with parameters $\vec{p}=\{5000, 1.0, 0.5\}$. The solid curves are 122 ray paths used in the analysis to compute flow kernels. The dashed half circle represents the radius $r = 0.7\,{\rm R_\odot}$. Right: Forward (noiseless) travel times computed from ray kernels for each distance and latitude.}
 \label{fig:raypaths}
\end{figure}


We use a simple, single-cell, meridional-flow model first described by \citet{vanballoon1988} and \rev{later} utilized by \citet{dikpati1999}, among others. The parametric model is given by Equations~57\,--\,61 of \citet{vanballoon1988} and will not be reproduced here. \rev{It is computed in a polar $(r,\theta)$ meridional plane.} For our purposes, the meridional profile has effectively three free parameters, which will be denoted $p_1$, $p_2$, and $p_3$. $p_1$ controls the flow amplitude, while $p_2$ and $p_3$ control the latitudinal and  radial (depth)  dependence of the flow structure, respectively. The model provides two-dimensional flows in the radial and latitudinal directions  $\vec{v}(r,\theta) = v_\theta(r,\theta)\hat{\vec{\theta}}+v_r(r,\theta)\hat{\vec{r}}$ that satisfy mass conservation in the 2D domain: $\vec{\nabla}\cdot \rho\vec{v}=0$. \rev{The density is a function that scales as $\rho\sim r^{-1.52}$, similar to \citet{vanballoon1988}, but slightly modified to match Model S \citep{jcd1996} in the region of interest.} The input values of the three parameters are such that the poleward surface flow reverses direction at $r\approx 0.79\,{\rm R_\odot}$. We use a grid that has 150 points in latitude  and 100 points in radius, covering $\theta=\pm 90^\circ$ and from $r=0.68\,{\rm R_\odot}$ to $r={\rm R_\odot}$.

Ray kernels are computed for a set of latitudes and distances that sample the model relatively well (although by no means exhaustively). We consider ten  skip distances from $2^\circ$ to $45^\circ$. The central latitude range is $\pm 77^\circ$, resulting in a total of 122 ray kernels. Figure~\ref{fig:raypaths} shows the given circulation model with  all ray paths overplotted. The weaker radial flows of the model are not shown here.
 
Synthetic forward travel-time differences are then computed from the flow model and kernels, shown on the right of  Figure~\ref{fig:raypaths}. To these travel times, artificial, \rev{normally distributed} random noise is added at two different levels: \rev{${\cal N}(0,\sigma_1^2)$ and ${\cal N}(0,\sigma_2^2)$.} In the \textbf{low-noise case}, \rev{$\sigma_1=0.016$~seconds}  is about 2\,\% of the rms of the travel-time differences (0.8~seconds), and about 20\,\% in the \textbf{high-noise case} (\rev{$\sigma_2=0.16$~seconds).} These noise levels roughly correspond to typical \rev{meridional flow} measurements made over three years and one month, respectively \citep{braun2009}.

\subsubsection{SOLA Solution to the Problem}
\label{sec:sola}

\begin{figure}
  \centering
  \includegraphics[width=1\textwidth,clip=]{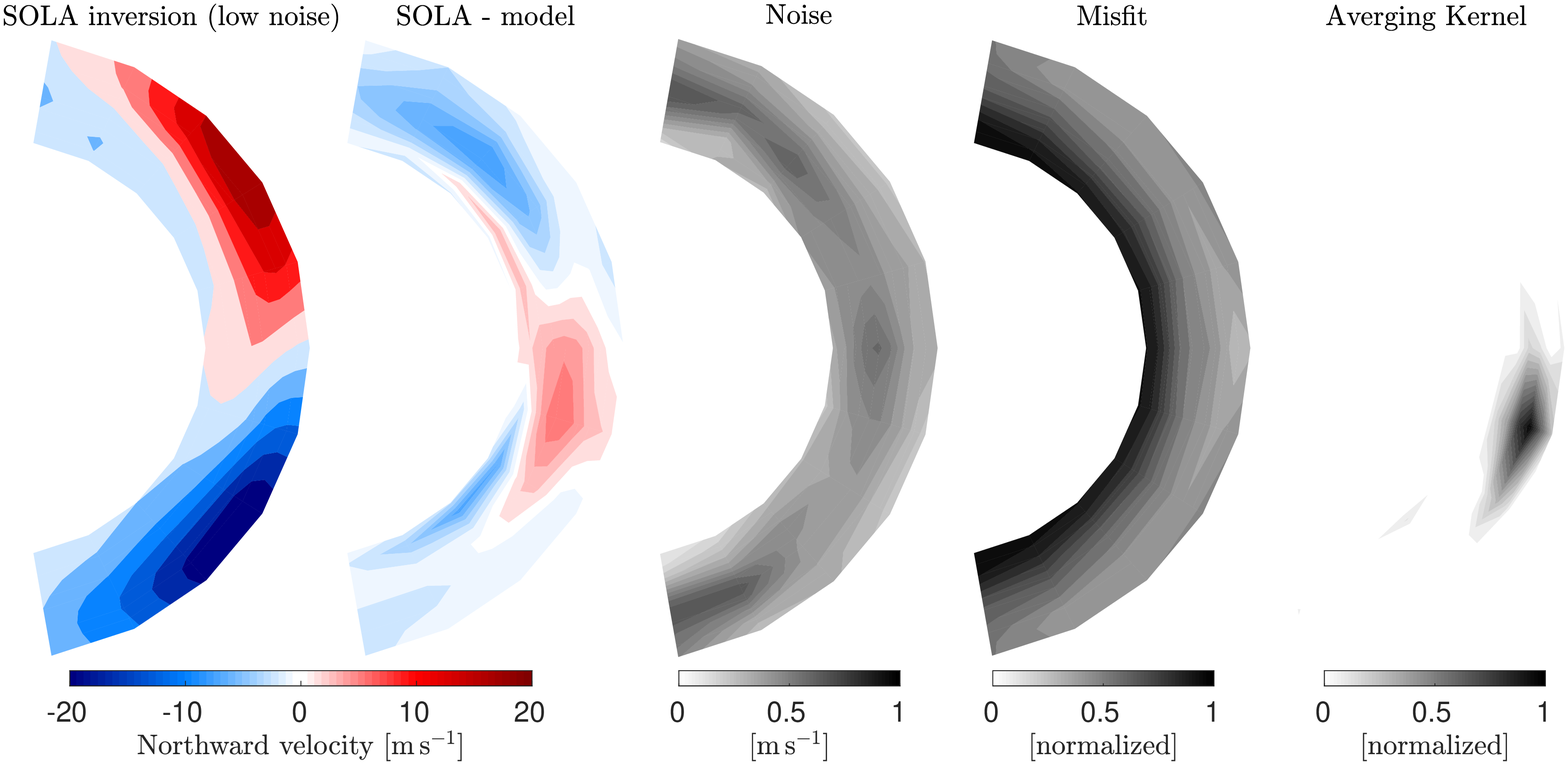}\\\vspace{.2cm}
  \includegraphics[width=1\textwidth,clip=]{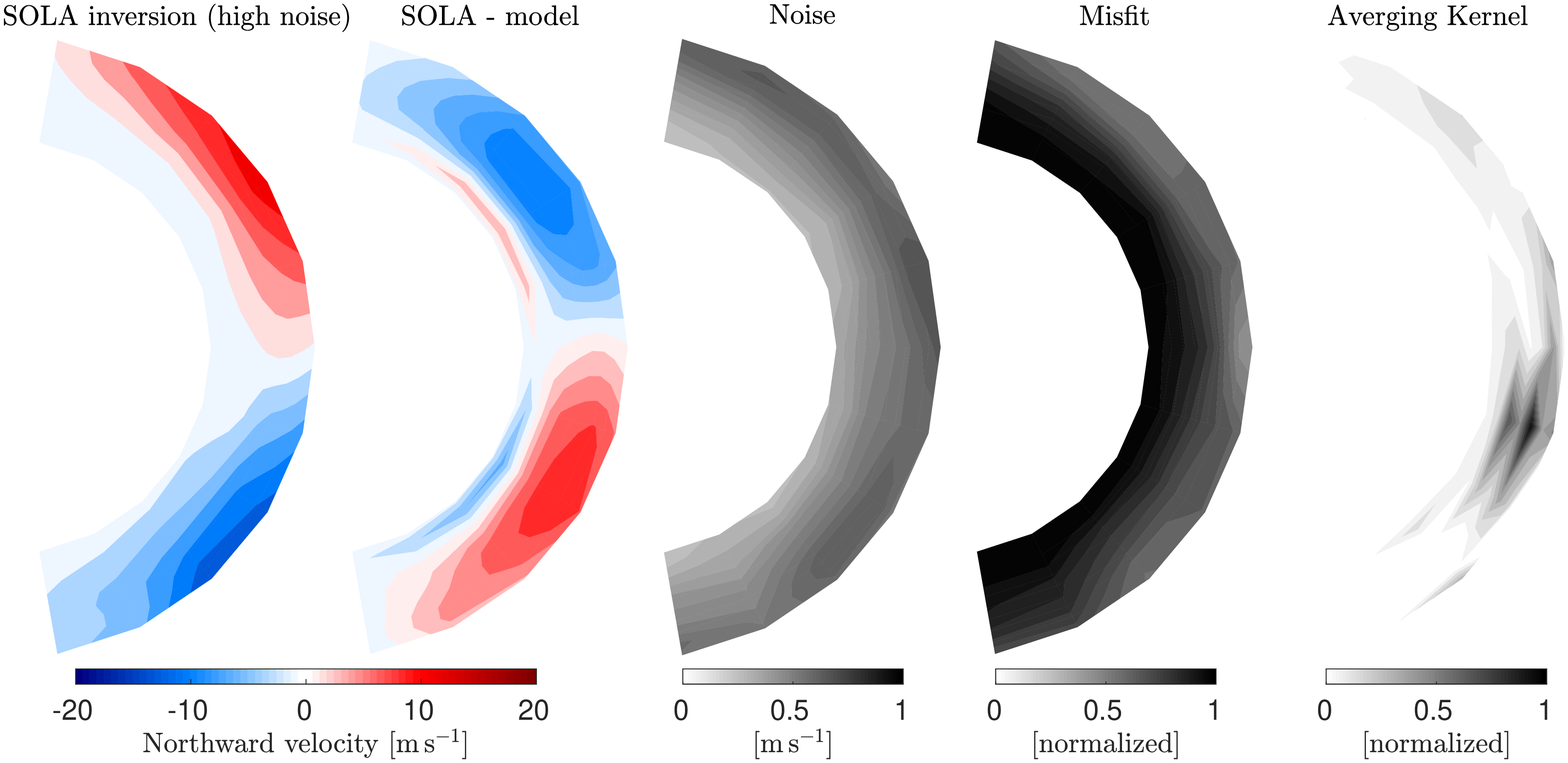}
  \caption{SOLA inversion results for travel times of different noise levels. The left two columns show the flows from the inversion and the difference with the known model. The middle column is the inferred noise at each inversion location. The fourth column is the misfit value at each inversion location. The last column is an example averaging kernel  from an inversion at a target location at $(r,\theta)=(0.9\,{\rm R_\odot},-15^\circ)$. \textit{Top row}: SOLA inversion for the low-noise case. The overall median noise is about ${\rm 0.4~m\,s^{-1}}$. \textit{Bottom row}: SOLA inversion for the high-noise case. The overall median noise is about ${\rm 0.5~m\,s^{-1}}$.}
  \label{fig:sola}
\end{figure}

We first demonstrate the standard inversion method described in Section~\ref{sec:typ}. It is the SOLA inversion applied by \citet{jackiewicz2015} and other recent studies. The synthetic travel-time differences are considered to be uncorrelated, and thus the noise-covariance matrix used in the inversion is diagonal. No mass-conserving constraint is imposed, and therefore it is hopeless to try to recover the small radial velocity in this inversion, which is about 10\,\% of the amplitude of the latitudinal flows.


The SOLA inversion estimates the velocities at specific \textit{target} locations. In this example, there are 110 target locations, 10 in depth and 11 in latitude. At each location, a 2D Gaussian target function was computed with a full-width-half-maximum (FWHM) in the radial direction of $0.08\,{\rm R_\odot}$ and in the latitudinal direction of $10^\circ$. The target function replaces the unrealistic $\delta$ function given in Equation~\ref{eq:cost}, and it gives a measure of the spatial resolution of the inversion results.

The results of the SOLA inversion are shown in Figure~\ref{fig:sola} after inverting the low-noise and the high-noise travel times. To aid in comparison with the known model, the retrieved flows at the 110 spatial locations have been interpolated onto the model grid. The recovered flows generally follow the pattern of the model. The deeper return flow is not reliably found in either case. Since the SOLA inversion always returns a flow pattern that is a smoothed version of the real one \citep[see][]{jackiewicz2012,svanda2012}, the amplitude is underestimated. On average, the underestimation is about $2~{\rm m\,s^{-1}}$ in the low-noise case, about $4.5~{\rm m\,s^{-1}}$ in the high-noise case, but in some locations up to  $10~{\rm m\,s^{-1}}$

The inferred noise is too small and not consistent with the errors. Specifically, the retrieved velocity is $\approx 10\sigma$ away from the input in the high-noise case. In other words, if the true answer were not known and we surmised that our result is within 1 or 2 $\sigma$ away from the truth, we  would make an error of one order of magnitude.  Another feature of the inversions reveals the expected less-localized averaging kernel for the case of the noisier travel times. Note that one can tune the trade-off parameters to obtain different results (smoother/less noisy, more localized/noiser, \textit{etc.}), making the interpretation of the validity of the inferences challenging.

\subsubsection{Probabilistic Solution to the Problem: Parameter Posteriors}

\begin{figure}
  \centering
  \includegraphics[width=1\textwidth,clip=]{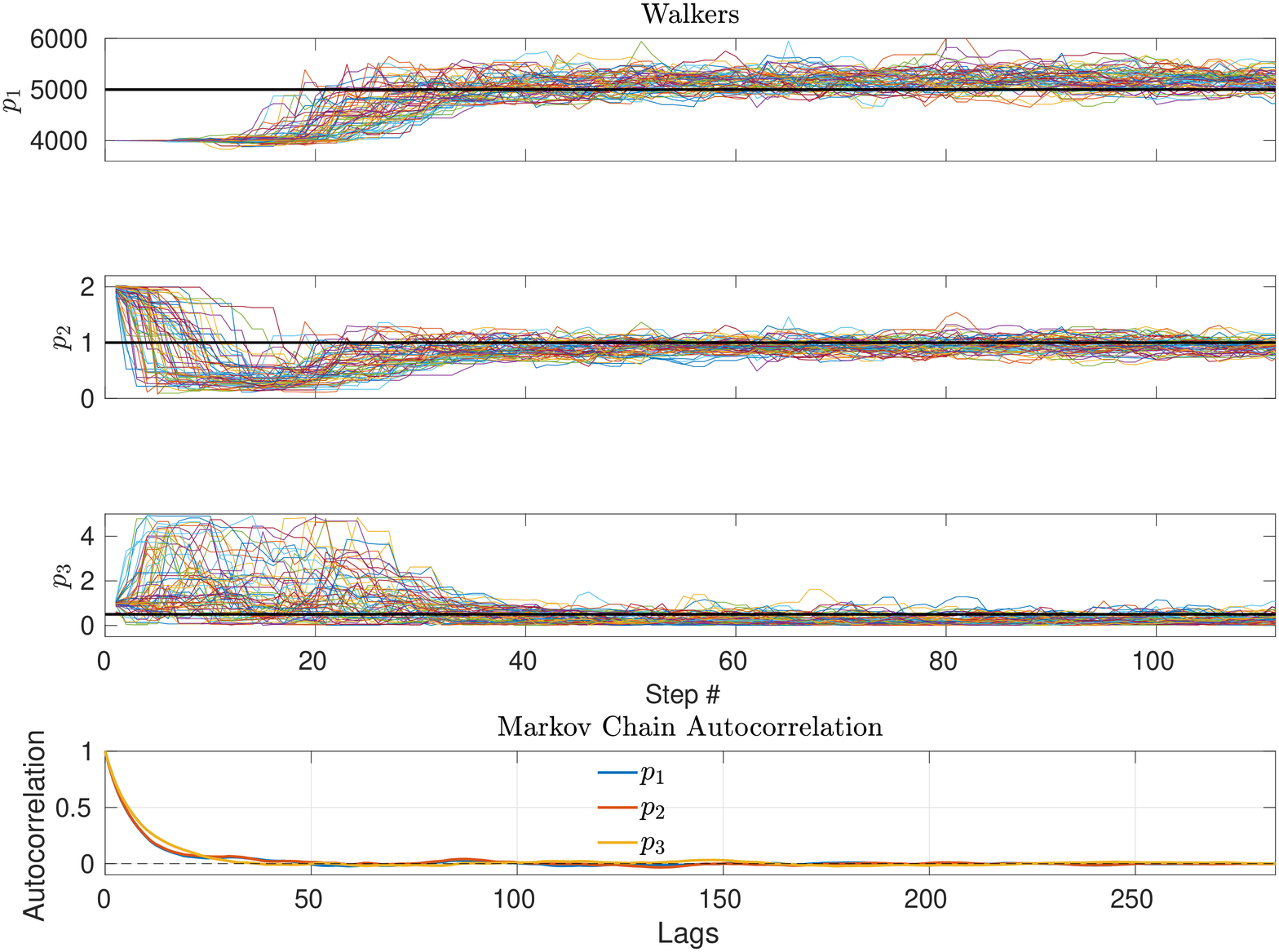}
   \caption{Probabilistic inversion diagnostics. The top three panels show how each of the 60 walkers of each parameter traverses parameter space. Each walker is a different color. The black horizontal lines are the input (known) values. The $y$-intercepts are the starting values of the walkers. Only the first 1/3 of the steps in the run are shown. The bottom panel shows the autocorrelation of each parameter's walkers as a function of the step lags. The burn-in phase was discarded before the calculation. The effective sample size is 1168.}
  \label{fig:post}
\end{figure}

Before showing the results of the Bayesian  MCMC inversion in a standard way, it is important to explore the output at the level of the walkers and the multidimensional PDF of the parameters.  In this example, the total number of steps (iterations) was chosen to be $10^5$. Each of the three free parameters was assigned 60 walkers (chains). The sampling of each walker was every five steps, which is a ``thinning'' procedure, whereby only the fifth step is stored. The PDF was therefore sampled $10^5/(60\times 5)=333$ times per walker. The choice of the standard deviation of  Gaussian likelihood function  is $\sigma=0.5$~second. Since the measurements are assumed to be uncorrelated, $(\mvec{\Sigma})_{ij} =\sigma_i^2\delta_{ij}$,  the likelihood function in Equation~\ref{eq:like} reduces to
\begin{equation}
  L(\vec{d}|\vec{m},\sigma_i) = \sum_i \frac{1}{\sigma_i\sqrt{2\pi}}\exp\left[-\frac{1}{2}\left(\frac{d_i-g(m_i)}{\sigma_i}\right)^2\right].
\end{equation}
The priors are taken as flat and rather wide, for demonstration purposes, as if we did not have a good idea of their values. \rev{These are known as ``uninformative'' priors.}

Figure~\ref{fig:post} shows the time series of all of the walkers during the run. Upon inspection,  the first thing to point out is that the initial  $\approx~30-40$ steps are when the sampling ``burns-in.'' This essentially means that the chains take a few steps to wander towards and reach a high-probability region, since the starting values typically might be far from such regions, as in this example (by choice). There is endless debate about burn-in validity in the literature \citep[\textit{e.g.}][Section~7]{hogg2018} into which we will not delve. In any case, the walker behavior is acceptable, in that once burnt-in, the space of the PDF is fully explored. The acceptance rate of the proposed steps is about 30\,\% -- a good value for MCMC algorithms.

Sample draws can be  correlated in MCMC algorithms due to noise or other factors. If each draw were completely independent, then the variance would decrease as more and more samples are drawn.  It is critical to know if independent samples are drawn from the PDF so that the parameter estimation is not biased, and reliable estimates of the mean/median and variance can be computed. The standard way of determining this is by calculating the autocorrelation of the walkers of each parameter. When and if the autocorrelation approaches zero, one can be confident that the walkers ``lose their memory'' of where they started and reach some state of equilibrium. The autocorrelation can be computed empirically from the time series of walkers, and it is shown in Figure~\ref{fig:post} in the right panel. In this case, the rate of convergence is quite rapid compared to the length of the run, and even fewer total steps could have been chosen. The effective sample size (ESS) is another concept to understand how many independent samples were drawn in the walker time series, and it is found from the autorcorrelation \citep{sokal1997}. In this example, the ESS is 1168. For a standard deviation $\sigma$ of the PDF of a given parameter, the  Monte Carlo standard error goes as $\sigma/\sqrt{\rm ESS}$. This means that we are  able to measure the median of a parameter with about a 3\,\% error compared to the overall uncertainty $\sigma$.

\begin{figure}
  \centering
  \includegraphics[width=1\textwidth,clip=]{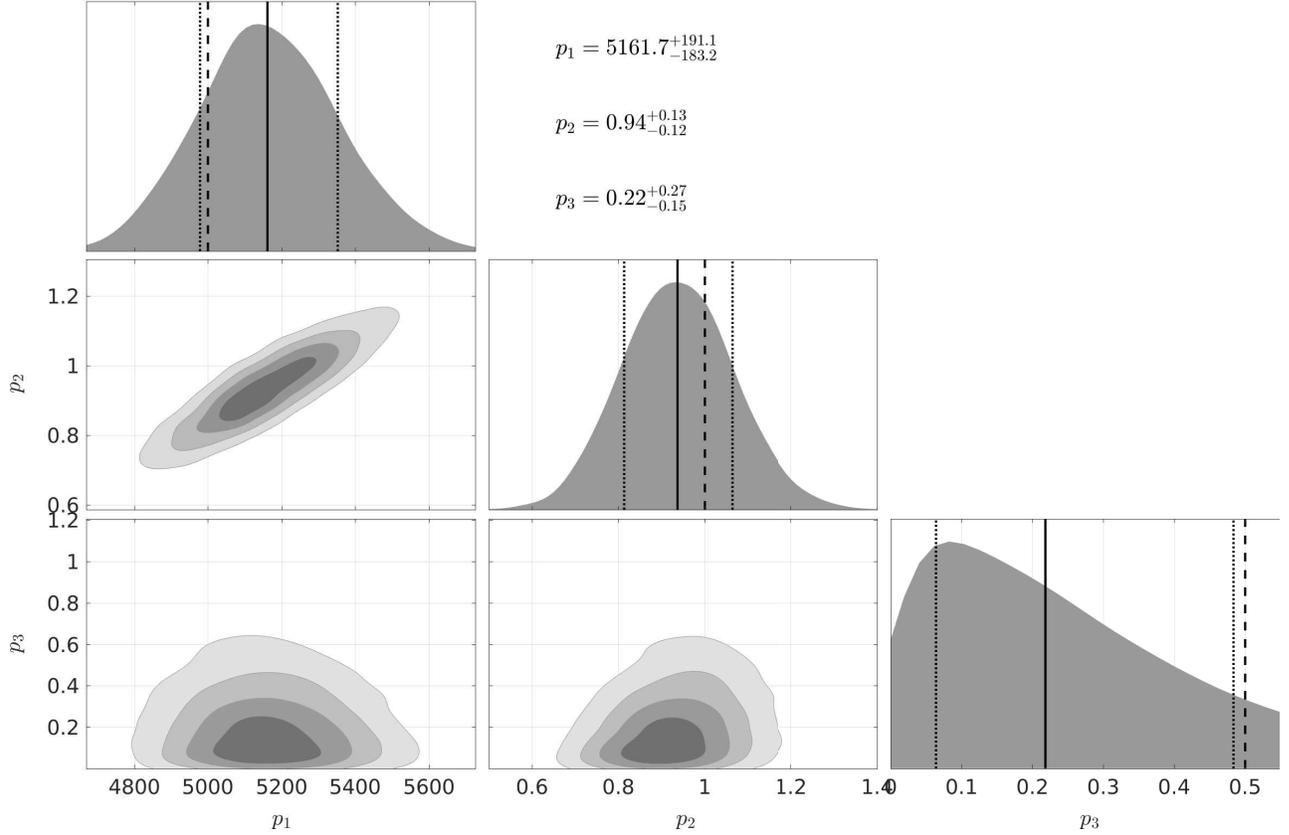}
  \caption{Corner plot showing the marginalized PDFs of the 3 parameters. The marginalized distribution for each parameter independently is shown in the histograms along the diagonal, and  the marginalized 2-D distributions as contour plots in the other panels. For each 1D histogram, the median of the PDF is the solid black line, and the \rev{dotted} lines enclose the 68\,\% confidence interval. The numerical values are given at the top.  The \rev{dashed black} lines are the known input parameter values. The contour levels of the 2D joint probability densities are at 20\,\%, 40\,\%, 60\,\%, and 80\,\% confidence intervals.}
  \label{fig:corner}
\end{figure}

The PDF in this example is three-dimensional. The ``corner'' plot matrix in Figure~\ref{fig:corner} shows histograms of all of the one- and two-dimensional projections of the PDF  of the parameters. The marginalized 1D PDFs are along the diagonal, and correlations between parameters are given in the off-diagonal elements (marginalized 2D PDFs). In this case, the PDFs are  not multimodal, which can be an indication that the model is parametrized well. This should not be surprising since the input model follows the same parametrization as  the forward model.

The power of the projected model parameter PDFs in  Figure~\ref{fig:corner} is that one immediately sees the distribution widths, as well as any correlations between model parameters. In this example, the PDFs bracket the known input value of the parameters within the 16 and 84 percentiles, except  parameter $p_3$, which is just beyond that range. This parameter also has the least Gaussian PDF.



\subsubsection{Probabilistic Solution to the Problem:  Model Space}

\begin{figure}
  \centering
  \includegraphics[width=1\textwidth,clip=]{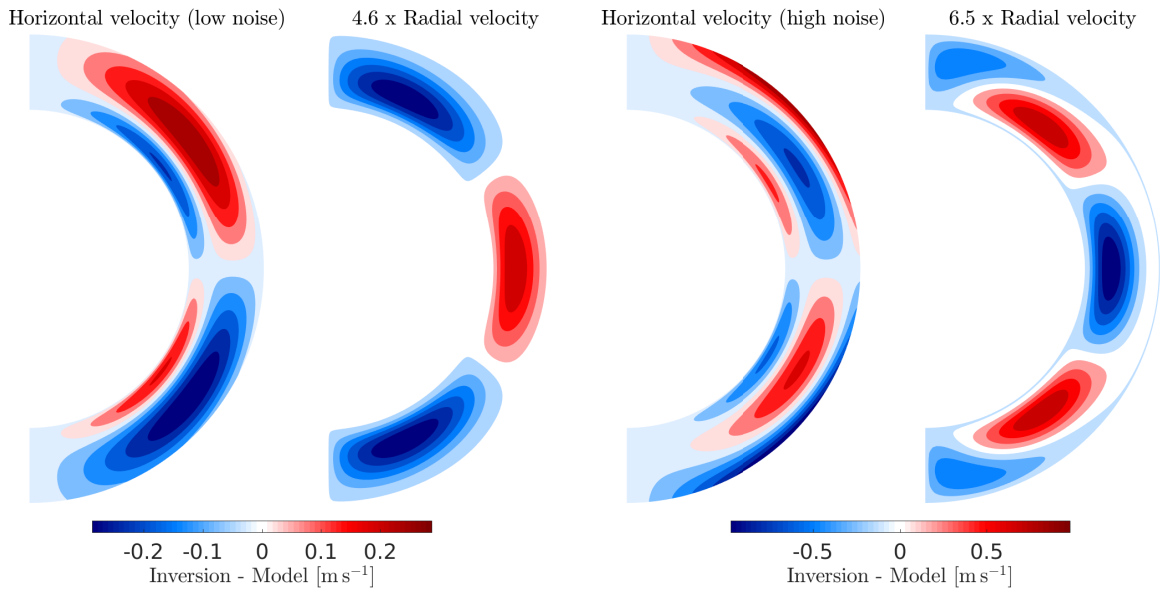}
  \caption{Results for both components of the meridional circulation using the probabilistic inversion. The inferred flows use the  median of the parameter PDFs. The panels show the difference of the two flow components inferred from the inversion with the model, for both levels of noise. The color scale extends to the limits of the data in each panel. The factor necessary to multiply the lower-amplitude radial velocities to achieve this is shown at the top.}
  \label{fig:mcmc_flows}
\end{figure}

The median of the PDFs of the inversion for the three parameters are used to generate a flow circulation profile. This provides a way to visualize the results in the space of the model, similar to what was shown earlier for SOLA. The resulting profiles  are very comparable to the input model, so much so that in Figure~\ref{fig:mcmc_flows} only the differences with the model are shown. The differences are significantly less than the SOLA example described in Section~\ref{sec:sola}. In the low-noise example, the inversion very slightly overestimate the poleward flow amplitude (parameter $p_1$), and therefore the equatorward flow is weakly underestimated. This affects the radial-velocity differences in the manner shown. In the case of noisier travel times, $p_1$ is again overestimated, but the other two parameters are slightly underestimated, leading to some small-scale differences in the relative flows.







\subsubsection{Probabilistic Solution to the Problem: Data Space}

\begin{figure}
  \centering
  \includegraphics[width=1\textwidth,clip=]{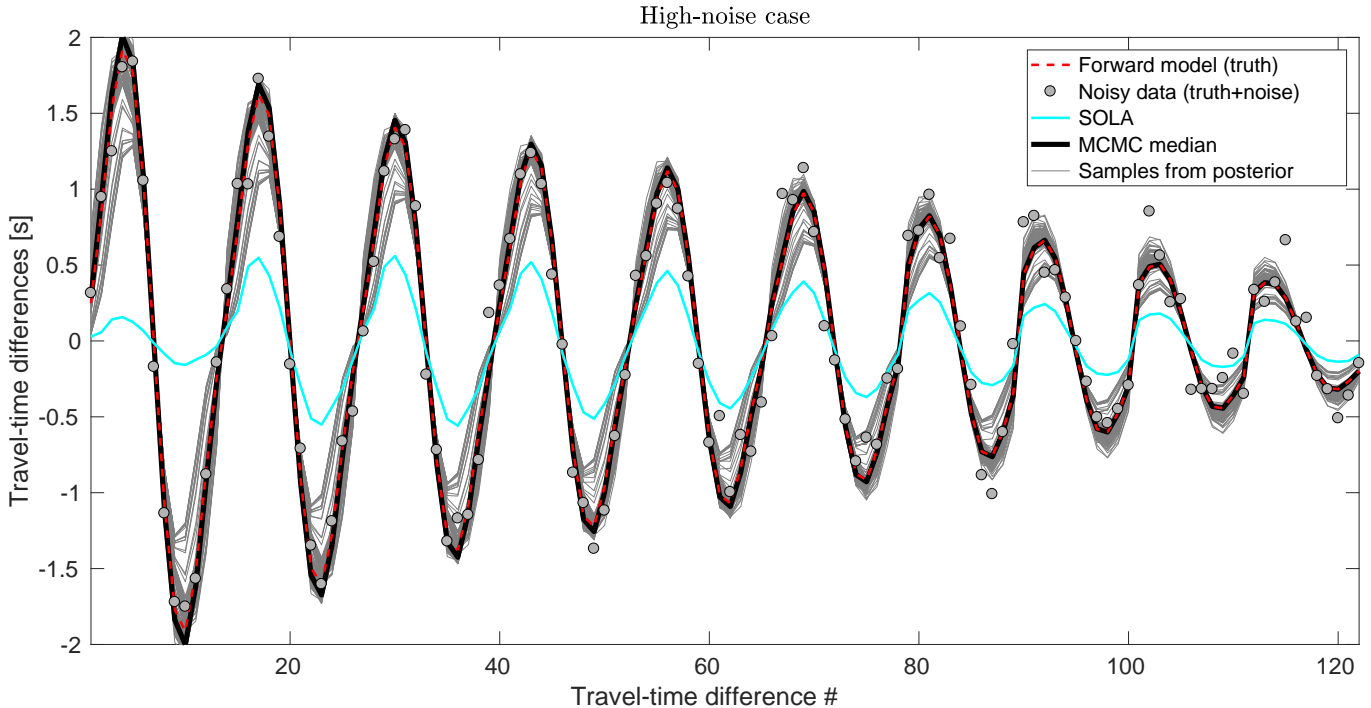}
   \caption{Inversion  solutions in the data (travel-time) space for the high-noise case. The dashed red line shows the input travel times, while the filled gray circles represent the random noise addition. The forward-modeled travel times from the SOLA inversion result are in cyan. The thick black line is computed from the median of the parameter PDF. Finally, the thin  gray lines are forward travel times computed from 100 realizations of the posterior distribution.  The travel times are plotted such that each ``oscillation'' is a different skip distance (smaller to larger from left to right), and the points within each oscillation correspond to each latitude \citep[see][for similar plots]{boning2017a}.}
  \label{fig:merid_data}
\end{figure}

The MCMC method allows for a visualization of the results in data space too, much more naturally and quickly than the SOLA method.
Figure~\ref{fig:merid_data} presents  the data-space solutions for both inversion methods using noisier travel times. One immediately sees the manifestation of the underestimated velocity in the SOLA inversion in the smaller-amplitude travel times. The near-surface region (left of the figure at smaller skip distances) is particularly evident. On  the other hand, the travel times generated from the median of the PDF are highly consistent with the input ones. Also shown are 100 random realizations of the PDF, which quickly gives a picture of the statistical uncertainties in the data space.

\subsection{Supergranulation in the Born theory}

The Sun's supergranulation is an important component of near-surface convection-zone dynamics and a strong source of advection of  magnetic fields. While its sub-surface flow structure has not yet been faithfully determined by helioseismology, it may have a simple enough form to be parametrized by a model. It is therefore another suitable test case for a probabilistic inversion. For this example, we consider five new aspects that add complexity and richness to the demonstration:
\begin{itemize}
\item The problem is set up in three dimensions (rather than two);
\item Born-sensitivity kernels are used (instead of ray kernels);
\item The observations are from a  3D numerical simulation with stochastic, realistic noise properties (not synthetic forward-modeled observations);
\item A proper noise covariance matrix is computed and used in the inversions (not just diagonal variances); and
\item The supergranule model in the simulation is \textit{different} from the model and parameters used to estimate the PDF.
\end{itemize}
Regarding the last point, this means that the ``true'' values of the parameters used to simulate the supergranule are essentially unknown, unlike the meridional-flow example where the input $p_i$ could be directly compared to the posteriors. We briefly describe the problem setup before studying the results.

\subsubsection{The Models}

The supergranulation model is taken from \citet{dombroski2013}. In that work, realistic wave propagation using the SPARC code \citep{hanasoge2006} was simulated through a single, kinematic supergranule flow pattern to quantify the effects on seismic waves. The mass-conserving flow structure was modeled using seven  parameters. Two control the horizontal extent of the divergent flow, three control the depth-dependence and strength of the outflow, and two more parameters control the depth dependence of the boundary inflow.

The \rev{model supergranule} has a radial extent of about $30$\,Mm at the surface, where the maximum horizontal speed and the vertical speed is $250\,{\rm m\,s^{-1}}$ and $20\,{\rm m\,s^{-1}}$, respectively. The outflow switches to an inflow at a depth of  $\approx 10$\,Mm. The maximum vertical speed is about $100\,{\rm m\,s^{-1}}$, peaked around $4$\,Mm below the photosphere. For purposes later, this will be referred to as the reference (``ref'') model.

As a notable aside, while this model is reasonable and at least consistent with surface observations \citep{duvall2010,rieutord2010}, supergranulation has proven very difficult to fully understand. There are even questions about whether models that have separable flows (in horizontal and vertical directions) are appropriate for supergranulation \citep{ferret2019,druv2019}. Addressing such issues is beyond the scope of this article.

The model that we use in the probabilistic inversion is instead from \citet{duvall2012}. Also employing a separable, mass-conserving flow, this model has five free parameters. In fact, two  of the parameters are equivalent between the models, those that control the horizontal diverging flow as
\begin{equation}
  \vec{g}(r) = J_1(kr)\exp\left(-r/R\right)\hat{\vec{r}},
\end{equation}
where $J_1$ is an order-one Bessel function, $k$ is a wavenumber, and $R$ represents a decay length in the distance coordinate from the origin [$r$]. The values from \citet{dombroski2013} are $k=2\pi/30\,{\rm rad\,Mm^{-1}}$ and $R=15$\,Mm, identical to those in \citet{duvall2012}.  In addition, this model has a Gaussian depth-dependence of the velocities, determined by three additional parameters: a peak amplitude [$v_0$], a peak flow location [$z_0$], and  a Gaussian width [$\sigma_z$], leading to the function
\begin{equation}
  u(z) = \frac{v_0}{k}\exp\left(-\frac{(z-z_0)^2}{2\sigma_z^2}\right).
\end{equation}
Once $\vec{g}$ is computed, the model vertical flows are constructed first as $v_z(r,z) = u(z)\vec{\nabla_{\rm h}}\cdot\vec{g}$. Then the horizontal flows are $\vec{v_{\rm h}}(r,z)=-f(z)\vec{g}(r)$, where $f$ is obtained from applying the continuity equation. We compute this model in three spatial Cartesian dimensions $(x,y,z)$ for illustration sake, even though it is axisymmetric and the problem can be solved in only two. This will be referred to as the ``trial'' model.

Apart from the two common free parameters, each model is derived differently enough that the priors on the other three free parameters are not well known. We take uniform priors that are kept identical in each of the probabilistic inversions discussed below. The priors are given in Table~\ref{tab:priors}.

\begin{table}
\caption{Table of (uniform) priors for the supergranulation probabilistic inversions using the ``trial'' model. They are ordered according to how the results are presented.}
\label{tab:priors}
\begin{tabular*}{\textwidth}{@{\extracolsep{\fill}} ccccl}     
  \hline                   
Parameter & Label & Min.  & Max. & Unit \\\hline
$v_0$ & $p_1$ & 50  & 400 & ${\rm m\,s^{-1}}$ \\
  $\sigma_z$ & $p_2$ & 0  & 12  & Mm \\
  $z_0$ & $p_3$ & 0 & -12\tablenotemark{a} & Mm \\
  $k$ & $p_4$ & 20  & 40 & ${\rm rad\,Mm^{-1}}$ \\
  $R$ & $p_5$ & 10  & 25 & Mm \\
  \hline
\end{tabular*}
\tablenotetext{a}{Negative values denote sub-surface}
\end{table}

\subsubsection{Setup of the Problem}
\label{sec:setup}

Helioseismic  measurements were computed from the numerical simulation using a time series of the vertical velocity, which is sampled every 1 minute at 200~km above the model photosphere over a total of 24 hours. The horizontal spatial domain extends 100~Mm and is sampled every 1/3\,Mm. The vertical velocity is first filtered to isolate different  ridges (radial orders $n$), including the $f$-mode ($n_0$) and the first two acoustic-mode ridges ($n_1$, $n_2$) using standard methods \citep{braun2008b,gizon2009,degrave2014a}. Cross correlations were measured in center-to-annulus and center-to-quadrant geometries for 15 different travel distances, ranging from 6\,Mm to 20\,Mm. For each ridge and each distance, three  travel-time difference maps are computed (at the same spatial resolution) across 50~Mm of the simulation domain: ``out--in'' [$\delta\tau_{\rm oi}$], ``west--east'' [$\delta\tau_{\rm we}$], and ``north--south'' [$\delta\tau_{\rm ns}$]. Such geometries are sensitive to flows \citep{duvall1997}. This results in 135 unique travel-time maps, which are very comparable to the ones computed using helioseismic holography by \citet{dombroski2013}. Only a fraction of these measurements are used in the sample inversions.

\citet{dombroski2013} computed a second simulation without a background supergranule. We use these data (split into twelve two-hour cubes) to estimate the noise covariance in the travel times according to the noise model of \citet{gizon2004}. The exact same measurement procedure explained above is carried out on these cubes to determine the covariances ${\rm Cov}[\delta\tau_i,\delta\tau_j]$.

The linear forward problem \citep{gizon2002} in this example can be written
\begin{equation}
  \delta\tau^\alpha_i(x,y) = \iiint\vec{K}_i(x-x',y-y',z)\cdot\vec{v}^\alpha(x',y',z)\,\rmd x'\rmd y'\rmd z,
  \label{eq:bornforward}
\end{equation}
where flows are $\vec{v}$, the sensitivity kernels $\vec{K}_i$ are vector-valued, and each index $i$ corresponds to a given ridge, geometry, and travel distance. Born-approximation kernels are computed from \citet{birch2007} in a point-to-point fashion, and then averaged over annuli to be consistent with the travel-time geometries $({\rm oi, we, ns})$. The $\alpha$-superscripts refer to  the particular measurement source or supergranule model under consideration.

We can be precise about this example.  The SOLA inversion is looking to infer $\vec{v}^{\rm ref}$ in Equation~\ref{eq:bornforward} given the kernels and the measurements $\delta\tau^{\rm ref}$ of $v_z^{\rm ref}(x,y,z=200\,{\rm km})$, and so $\alpha={\rm ref}$. Let's call the estimate $\vec{v}^{\rm SOLA}$. The probabilistic inversion is using Equation~\ref{eq:bornforward} with $\vec{v}^{\rm trial}$ and the same kernels to forward compute $\delta\tau^{\rm trial}$, thus $\alpha={\rm trial}$ in that case. The $\delta\tau^{\rm trial}$  are used in the  computation of the likelihood function along with $\delta\tau^{\rm ref}$. In  model space, the probabilistic  inversion is seeking to estimate suitable values of  parameters such that $\vec{v}^{\rm trial}$  will resemble $\vec{v}^{\rm ref}$. Further details in the setup of the inversions are mentioned in Appendix~\ref{app:inv}.

\subsubsection{Results}
\label{sec:sgresults}

SOLA inversions in 3D are not very convenient to compare with the probabilistic inversions, neither in model space nor data space. The reason is that, typically, the flows in $(x,y)$  are inferred one depth at a time, and these depths are usually few. Furthermore, the prescribed ``resolution'' in both directions can vary from depth to depth. In this example, the SOLA inversions were carried out at three target depths $z_0=(0,-3,-4.5)$\,Mm. Each target depth had a different target width in the horizontal and vertical directions. The actual depth at which the inversion is most sensitive can also be distant from $z_0$ due to non-localized averaging kernels.  By contrast, and by construction, the  probabilistic inversions provide parameters that allow one to  estimate flows  over the full (or any) spatial domain.

To make meaningful comparisons between inversion results and the reference model, several steps need to be taken. Since the SOLA results are rather coarse in depth and smooth in the horizontal direction, we decide to adapt everything else to them. Firstly, for a given target depth, the reference model velocities are convolved  with the target function of the inversion  [the $\mvec{I}$ in Equation~\ref{eq:cost}, which in this case is not a $\delta$-function but a 3D Gaussian sphere]. This process is represented by Equation~\ref{eq:mtrue}, whereby the estimated flows are a smoothed version of the true ones: $\vec{v}^{\rm SOLA}=\mvec{R}\vec{v}^{\rm ref}$. The result is then integrated over depth, giving a 2D flow map that can be compared to the SOLA inferences. 


\begin{figure}
  \centerline{
    \includegraphics[width=.5\textwidth]{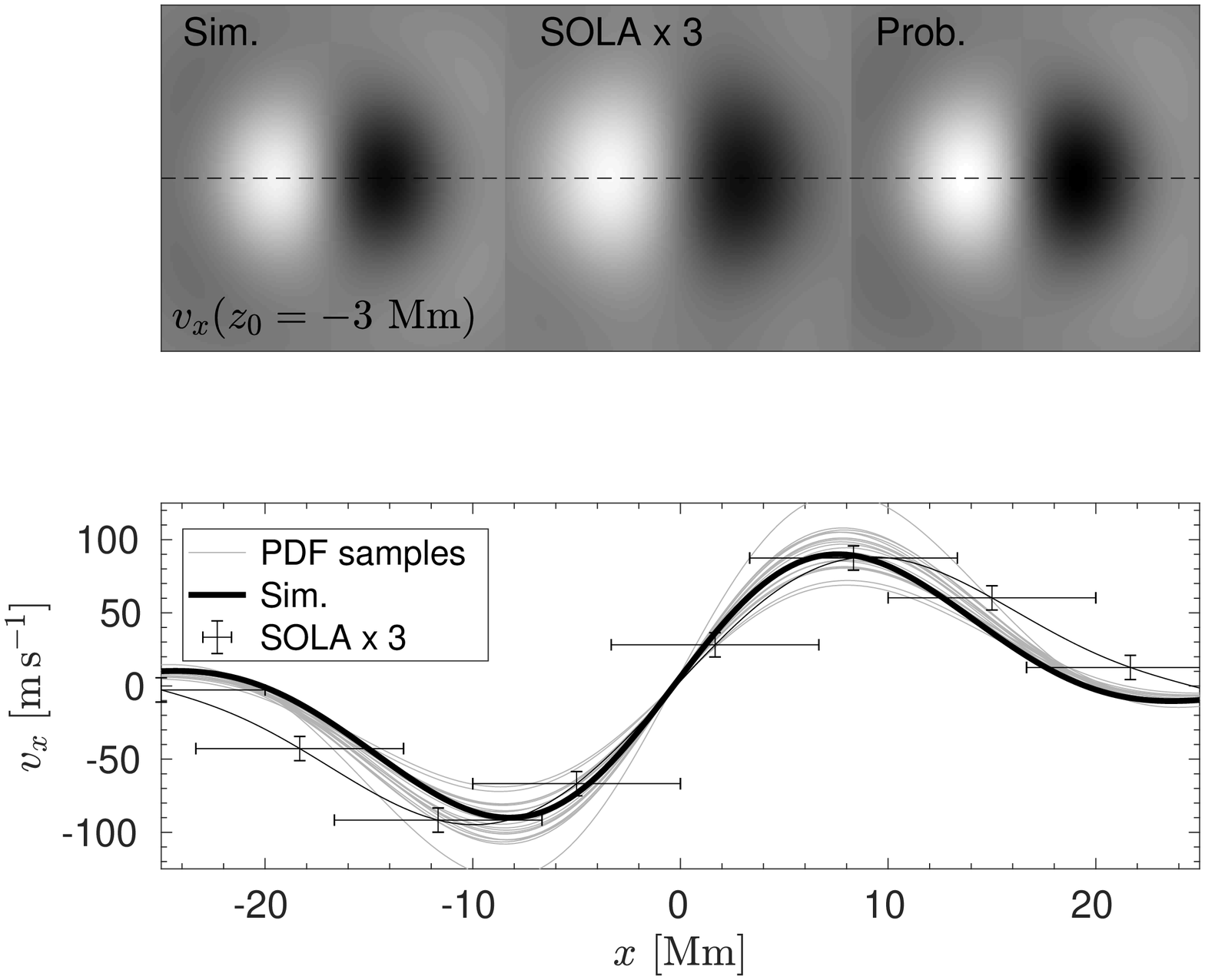}
    \includegraphics[width=.49\textwidth]{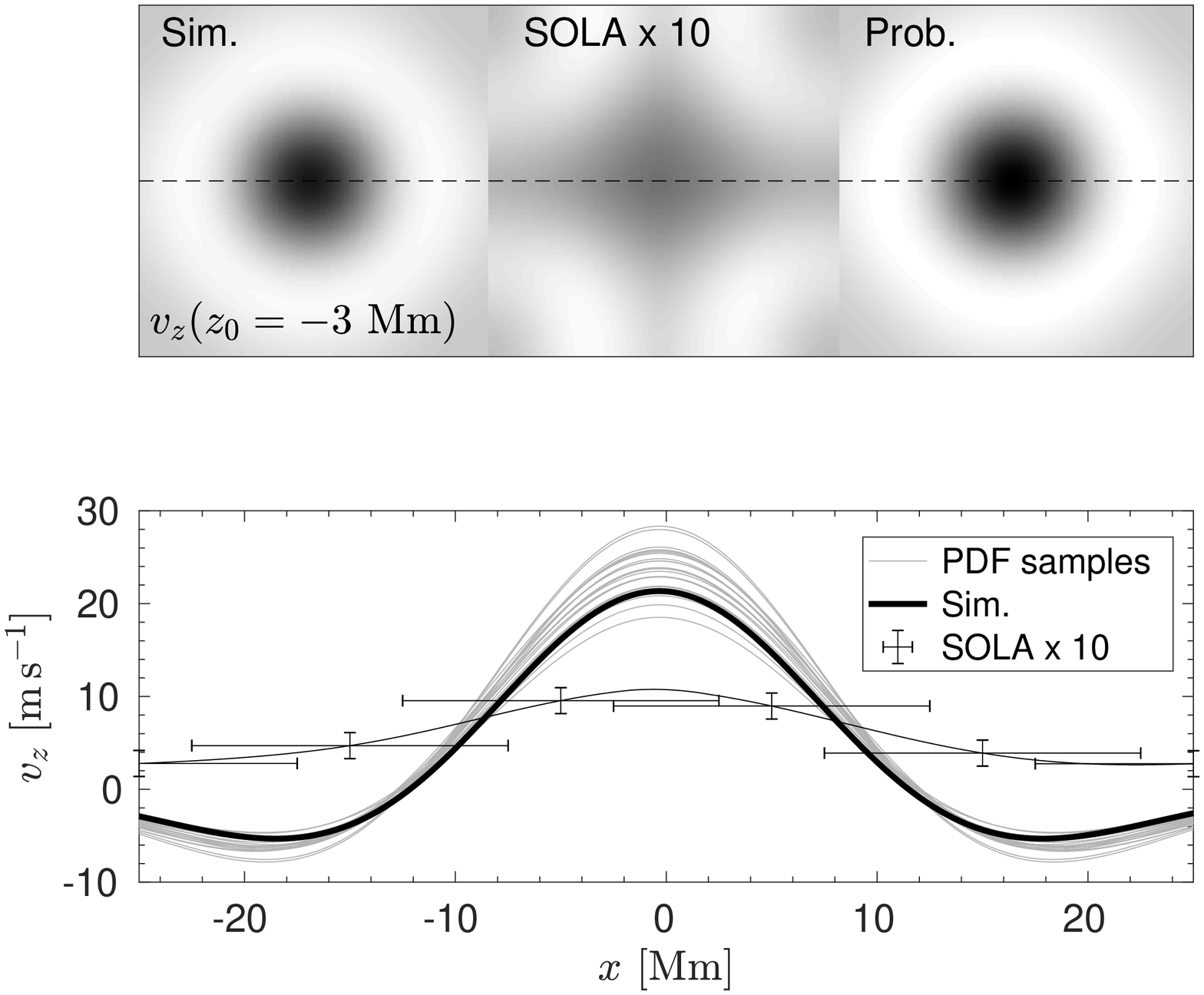}}
  \caption{Comparison of supergranulation inversion results at 3\,Mm beneath the model photosphere. The left (right) panels are for the $v_x$ ($v_z$) inversion. The top rows, from left to right, are the flow fields for the simulation, the SOLA inversion, and the probabilistic inversion computed from the median of the PDF.  Darker shading corresponds to positive velocities (to the right for $v_x$, and out of the page for $v_z$), and the scale is the same in each set. The bottom panels are cuts through the supergranule at $y=0$, shown by the dashed line in the top panels. Twenty random samples of the PDF are drawn and computed in the model space. A few representative points from the SOLA inversion are given with uncertainties.The SOLA velocities are scaled by the factor indicated.}
  \label{fig:v-3}
\end{figure}

For the probabilistic inversion, we use draws of the  model PDF parameters (median or otherwise) and compute the flow model $\vec{v}^{\rm trial}$ on the same spatial grid as $\vec{v}^{\rm ref}$ and the $\vec{v}^{\rm SOLA}$. It is also appropriately smoothed by the SOLA inversion target function and integrated over depth in the same manner. This process results in three sets of flow maps at three nominal target depths for three flow components, although we restrict comparisons to $v_x$ and $v_z$. Only results in model (velocity) space will be presented.

Figure~\ref{fig:v-3} shows a comparison of these two flow components at 3\,Mm beneath the photosphere, where the horizontally divergent structure is apparent.  In general, we find the SOLA inversions severely underestimate horizontal velocities (note the scaling factor),  while the probabilistic inversions weakly overestimate them.  The bottom panel of Figure~\ref{fig:v-3} shows a cut through the models at $y=0$. The noise in the SOLA inversion, even using covariance matrices, is highly underestimated.  The horizontal error bars show the FWHM of the target function, which were quite wide to get sensible results. On  the other hand, a random sample of the parameter PDF from the probabilistic inversion gives a reasonable spread of solutions in model space.

At the same depth, the inferences on the weaker vertical velocity are also shown in Figure~\ref{fig:v-3} on the right. In this case, the SOLA flow inferences are marginal at best. At inversions just below this depth, the SOLA flows are anticorrelated with the reference flows, as in Figure~\ref{fig:v-5} in Appendix~\ref{app:flows}. \citet{dombroski2013} found the same result in their inversions of this model, and they demonstrated that the culprit was the ``cross talk'' between vertical and horizontal flows that the sensitivity kernels, and inversions, are unable to disentangle. In our SOLA inversion, an explicit cross-talk term is included \citep{svanda2011}, and even still, the problem persists. The sensitivity kernels are not completely accurate. This can be verified by  comparing measured and forward-modeled travel-time differences, and as \citet{dombroski2013} showed (and we verified) there are anomalies in some of the travel-time maps. However, since both inversions use the same kernel functions, the relative comparsions are  meaningful.  Examples at other depths are given in Appendix~\ref{app:flows}.

Figure~\ref{fig:corner_sg} in Appendix~\ref{app:flows} shows the corner plot for the probabilistic inversion. The only ``known'' parameters are $p_4=R$ and $p_5=k$, so  comparisons between the input ones and inferred ones cannot be made due to the differing models. The probabilistic inversion overestimates the flow speeds, and is  mainly due to the estimation of the $p_2=\sigma_z$ and $p_3=z_0$ parameters (Section~\ref{sec:var} gives more evidence of this). These control  the location of the peak of the vertical-velocity profile  and its width. There are (at least) two reasons for the poor estimation of $p_2$ and $p_3$. The first, as the corner plot shows in the 2D marginalized PDFs, is that these two parameters are not highly correlated with the others, but more so with themselves. There must be some correlations due to the continuity-equation constraint, but it is a weak one. This could indicate a poor parametrization of this particular model for supergranulation\footnote{Indeed, the conclusions from earlier work using this model \citep{duvall2012,duvall2014} showed a very shallow supergranule that is quite different than what other studies have found in the literature. While the analysis may very well  be correct, more checks on the model need to be carried out, which is beyond the scope of this article}. The second reason has to do with the sensitivity functions used here. They have very little sensitivity below 8\,Mm, while the depth of the profile extends to a depth of about 12\,Mm. The likelihood is thus not informative, and the 1D PDF for parameter $p_3$ is not very Gaussian.


\subsubsection{Does Additional Data Bring New Information?}
\label{sec:var}

For researchers who have experience computing inversions in local helioseismology, it can be non-trivial to understand how the addition of extra observations will (positively or adversely) affect  the results. Indeed, a brief discussion regarding this point is  presented by \citet{dombroski2013} in their results section. For instance, consider  one set of measurements using particular seismic waves. Now, consider another set of measurements using the same seismic waves but where the only difference is different travel distances. Will including the second set with the first improve the inversion, just add unwanted noise, or improve the noise? Just doing this experiment may not  answer the question either, since the differences may be subtle, and SOLA or RLS inversions can be very sensitive to \rev{any outlier} measurement points.

The probabilistic inversion provides a way to study this question more quantitatively. We design a simple demonstration experiment, and leave a full analysis to another article. Eleven probabilistic inversions are computed, each one having different combinations or different numbers of  input data sets. Everything else is kept fixed. 

Two metrics are then calculated to assess the results. We compare the variance of the priors to the variance of the posteriors. Imagine the worst case scenario, when the variance is not reduced at all. This would imply that the addition of data has provided no new information on the model parameters. The goal of any inversion is to reduce the variance of the parameter estimation. The variance reduction metric is computed as
\begin{equation}
  \frac{{\rm var}\left[{\rm PDF}(\vec{m})\right]-{\rm var}\left[\rho(\vec{m})\right]}{{\rm var}\left[\rho(\vec{m})\right]}\times 100,
\end{equation}
where $\rho(\vec{m})$ is the prior distribution function (Equation~\ref{eq:bayes}) whose range of values is in Table~\ref{tab:priors}. The other metric is the simple correlation coefficient between the travel-time measurements and the forward measurements computed using the median of the PDF.

\begin{figure}
  \centering
  \includegraphics[width=\textwidth]{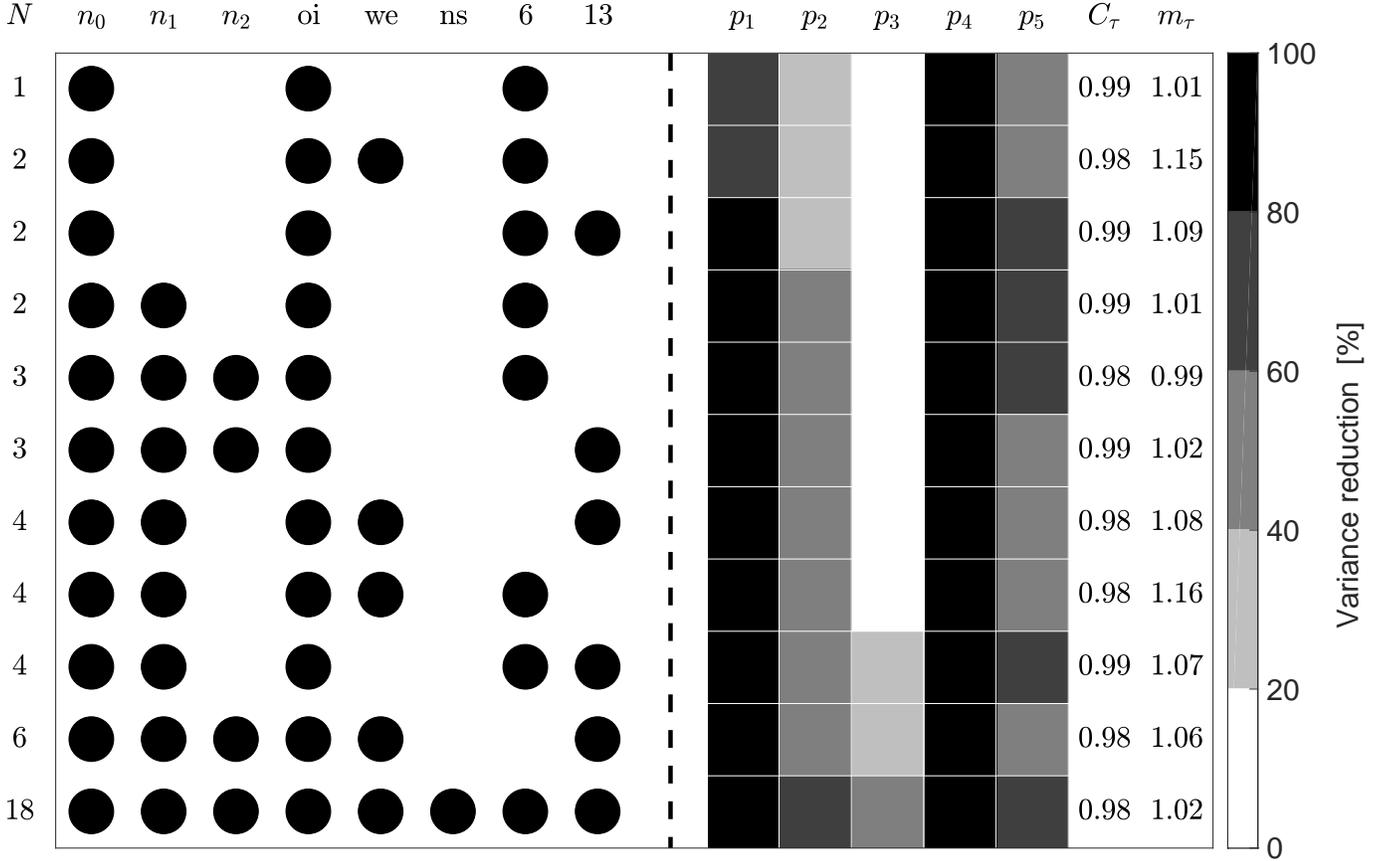}
  \caption{Metrics for 11 example probabilistic inversions. Each inversion comprises $N$ (left-most numbers) travel-time maps, consisting of the configuration shown by the next eight columns of the matrix. The $n_i$ denote the mode radial order, then the annulus geometry, and then the travel distances [Mm]. Filled circles indicate inclusion in the inversion. Beyond the vertical dashed line, the next five columns represent the variance reduction in the PDF of the parameters compared to the priors (given by the gray scale). The final two columns give the correlation coefficient $C_\tau$ between the inversion results and the measurements in data space.  Also given is the slope  [$m_\tau$] of a fit to the correlated data sets. }
  \label{fig:matrix}
\end{figure}

The results are provided in Figure~\ref{fig:matrix}. To understand what is shown, consider, for example, the second row of the matrix. $N=2$ means there are two travel-time maps used, $\delta\tau_{\rm oi}$ and $\delta\tau_{\rm we}$ for the $f$-mode at a travel distance of 6\,Mm. The black \rev{circles} for these quantities are filled. To the right of the dashed line, the variance reduction of the five parameters (as a percentage) are given by the gray scale. To be specific, the values for $p_1$ through $p_5$ in row 2 are $[74.7, 34.2, 1.6, 97.0, 55.5]\,\%$. The data have not provided much information on $p_3$ at all, as suspected. After that, $C_\tau$ is the correlation coefficient, and $m_\tau$ is the slope of a simple linear fit between the travel-time vectors. In almost all trials, the inferred data have larger amplitudes ($m_\tau\gtrsim 1$).

The third row is an inversion with only one change: the $\delta\tau_{\rm we}$ measurements are removed and an extra travel distance is added. The precise values of the variance reduction are now  $[82.5,   27.6,    0.4,   97.2,   68.2]\,\%$. The first parameter has gone from dark gray to black, over the 80\,\% level,  $p_2$ and $p_3$ are marginally worse, and $p_5$ is marginally better. Finally, the fourth row also uses two sets of measurements, with only one annulus geometry and one distance, but now two ridges ($n_0$ and $n_1$). This results in a better variance reduction of $p_2$ than the other cases, and brings the slope closer to one. One might conclude, for this scenario, that given a very limited number of measurements, it is best to use more ridges than adding distances or anything else.

One can continue this way for the other cases to find interesting trends. Inspecting the matrix as a whole, a few things stand out. $p_1$ and $p_4$ are the best ``resolved'' parameters, and $p_2$ and $p_3$ are the least. A quick glance at the PDFs in the corner plot in Figure~\ref{fig:corner_sg} confirms this. The last row in the matrix is from an inversion using 18 different travel-time maps, and the variance reduction for $p_2$ and $p_3$ are 62\,\% and 40\,\%, respectively, the best in the set. The correlation between maps is consistently high, and the slope fluctuates a bit, but is overall acceptable.

To answer the question posed in this subsection -- yes, at least in this particular example. While the addition of new data might not visually and qualitatively improve the comparison in data space or model space (as evidenced by the  unchanging correlations), the variance, or uncertainties of the parameters, generally does improve.

In principle, such an analysis could also be achieved from the SOLA inversions, but it would be  more cumbersome. Minimizing the cost function properly, \textit{i.e.} calculating a good averaging kernel,  becomes much more difficult with fewer and fewer kernels. Then the trade-off parameters change and some of the results would not be sensible. However, in the probabilistic framework, this is entirely reasonable and instructive.

\section{Discussion}




The previous sections have contrasted two linear methods for interpreting helioseismic measurements.  For comparison, we label the class of inversions similar to SOLA as  Method 1, and the class of statistical and probabilistic inverions as Method 2.  There are several similarities between these two approaches. Both methods require a type of forward equation relating the unknowns and measurements. Several such equations are provided (Equations~\ref{eq:gm}, \ref{eq:forward}, \ref{eq:bornforward}).  Both methods can be run numerically using parallelization when formulated appropriately.

The practical differences outnumber the similarities. Method 2 needs more than a forward equation; it requires a generative model that can be parametrized with a manageable number of parameters. \rev{Otherwise, the computational cost may become prohibitive.} It would be difficult to utilize Method 2 to make synoptic flow maps of the Sun that contain many different convective structures and size scales, as standard ``pipeline'' inversions now do for local helioseismic data \citep{zhao2012}. \rev{It would require hundreds of parameters, with very little prior information.} \rev{Similarly, there is no pre-determined form of the solution when Method 1 is used, and as such it cannot incorporate priors like Method 2.} Method 1 does not use the data until the last step, whereby it combines the measurements in an ``optimal'' way based on how the sensitivity functions were combined \rev{(it does use the noise covariance, however, in the computation of the large matrix)}.  Method 1 requires ways to deal with computing the inverse of a large, usually ill-conditioned matrix. Method 2 provides a statistical interpretation of the solution, while Method 1 is forced to provide a ``best model.''

Beyond similarities and differences, inversion methods need to be validated. Numerous helioseismic studies over the past decade have employed numerical models for validation purposes. This is a powerful strategy, since one can quantitatively test inversion results on the known answer from the model. The results of these studies provide very consistent  conclusions. On the one hand, there are those that use  Method 1 and measurements of (non-magnetic) numerical models that do not have realistic noise, although usually some form of noise is added to the measurements after the fact. The findings are generally encouraging  \citep[\textit{e.g.} example 1 in this article;][]{hartlep2013,jackiewicz2015,korda2019}. This would seem to  indicate rather persuasively that Method 1, as well as the sensitivity functions (either ray or Born), can be used to accurately solve problems. On the other hand, when more realistic simulation models are studied in the same way, the results are \textit{somewhat} in agreement near the surface, but quickly diverge below  $\approx 3$~Mm \citep[example 2 in this article;][]{zhao2007,dombroski2013,degrave2014a,degrave2018}. The solar-like realization noise in these models is a significant barrier, which brings skepticism to  \textit{any} inversion results using actual solar data and Method 1  \citep[as commented on by][]{braun2008,svanda2015,korda2019b}.

Despite heroic efforts and substantial progress, there unfortunately have not  been as many significant advancements as one would expect in our understanding of the Sun from explicit inversions of local helioseismic data \citep{gizon2010}. The two examples in this work are cases in point, where still no consensus has been established \rev{regarding supergranulation and meridional circulation} \citep{giles1997,zhao2013,rajaguru2015,liang2015a,jackiewicz2015,duvall2012,hathaway2012,greer2016}.

Indeed, most of the fundamental breakthroughs in local helioseismology have come from the observations alone, rather than the formal interpretation of them. Examples include far-side imaging from acoustic holography and time distance \citep{lindsey2000,zhao2007b},  direct imaging of large-scale flows \citep{woodard2002}, acoustic absorption by sunspots \citep{braun1997}, flared-induced sunquakes \citep{kosovichev1998}, and the recent detection of solar Rossby waves using different  helioseismic measurement strategies \citep{loptien2018,hanasoge2019,proxauf2020}, among many others.

The potential issues that inhibit a full helioseismic analysis of certain outstanding problems include systematics and realization noise inherent in measurements, the theoretical treatment of seismic wave scattering from solar perturbations, and the inverse methods applied. There are many ways for dealing with each of these factors at various levels, and this work provides a possible avenue forward for exploration of the inversion component.

\section{Summary and Outlook}

In this article we described a probabilistic inversion scheme for time--distance helioseismology that  uses Bayesian statistics and Monte Carlo sampling. A few simple examples were carried out and compared with the commonly used SOLA technique. The examples used synthetic data where the known answer was the target of the inversions. Given that the input sets of measurements and sensitivity functions were rather minimal, the goal was not to solve these problems completely (\textit{i.e.} infer the flows as well as possible), but to demonstrate some of the strengths and weaknesses of these two approaches.

While the examples were highly idealized, the intercomparison consistently showed that the SOLA inversions systematically underestimate  the flow speeds and the noise levels, \rev{compared to the other method. This may not be too surprising given that Method 2 exploits a generative model with  relatively few parameters. It is not surprising either that the solutions using Method 2 are always smooth, since they are constructed as such. However,} the probabilistic inversions also crucially provide informative posterior probability distribution functions on the model parameters that are more consistent with the known answer. \rev{This was the case even when using uninformative priors.} One may  question the need to use \rev{Method 2} for (likely) highly linear problems like meridional circulation. In some of the example cases, however, the posteriors are not Gaussian, which could be a possible reason why SOLA or least-squares methods are not optimal. At the very least, the probabilistic  method  could be used to explore which helioseismic problems have complex PDFs and demonstrate why other inverse methods get trapped in local minima.

SOLA inversions can be tuned at some level to obtain different properties of the solution. A particular advantage of the probabilistic inversion scheme is that many realizations of the solution (in data space and/or model space) are computed automatically, allowing for a broad view of any particular model and its relative probability given the data.

Future work in this area should concentrate on developing  well-parametrized models of solar structures amenable to helioseismic investigation. For example, the recent meridional circulation model of \citet{liang2018} is much more flexible than the one presented here. Good models may also increase computational efficiency. Finally, one could imagine forward (generative) models that compute other observables than travel times, such as the more fundamental and information-laden cross correlations. This would be another move in the direction of full-waveform inversions.

While this work is focused on time--distance helioseismology, application to ring-diagram analysis, helioseismic
holography, or direct modeling is straightforward.  For those interested in similar applications, the affine-invariant ensemble MCMC algorithm has been made available in Python (\textsf{emcee}: \url{https://github.com/dfm/emcee}) and Matlab (\textsf{GWMCMC}: \url{https://github.com/grinsted/gwmcmc}) and can be adapted to many types of problems.




%
\acknowledgments
  This article is dedicated to the life and work of the late Michael J.~Thompson, who unselfishly taught me a signficant amount of inverse theory.  The author wishes to thank Aaron Birch, Shukur Kholikov, Aleczander Herczeg, and Jon Holtzman for fruitful discussions, as well as Doug Braun for making the simulation data publicly available. The article made use of the \textsf{GWMCMC} code written by Aslak Grinsted. Partial funding support is acknowledged from the National Science Foundation under Grant Number 1351311 and from NASA under Award Number 80NSSC18K0672.

  Appendix
 \appendix

 \section{Supergranulation Inversion Details}
 \label{app:inv}

 \begin{figure}
  \centering
  \includegraphics[width=\textwidth]{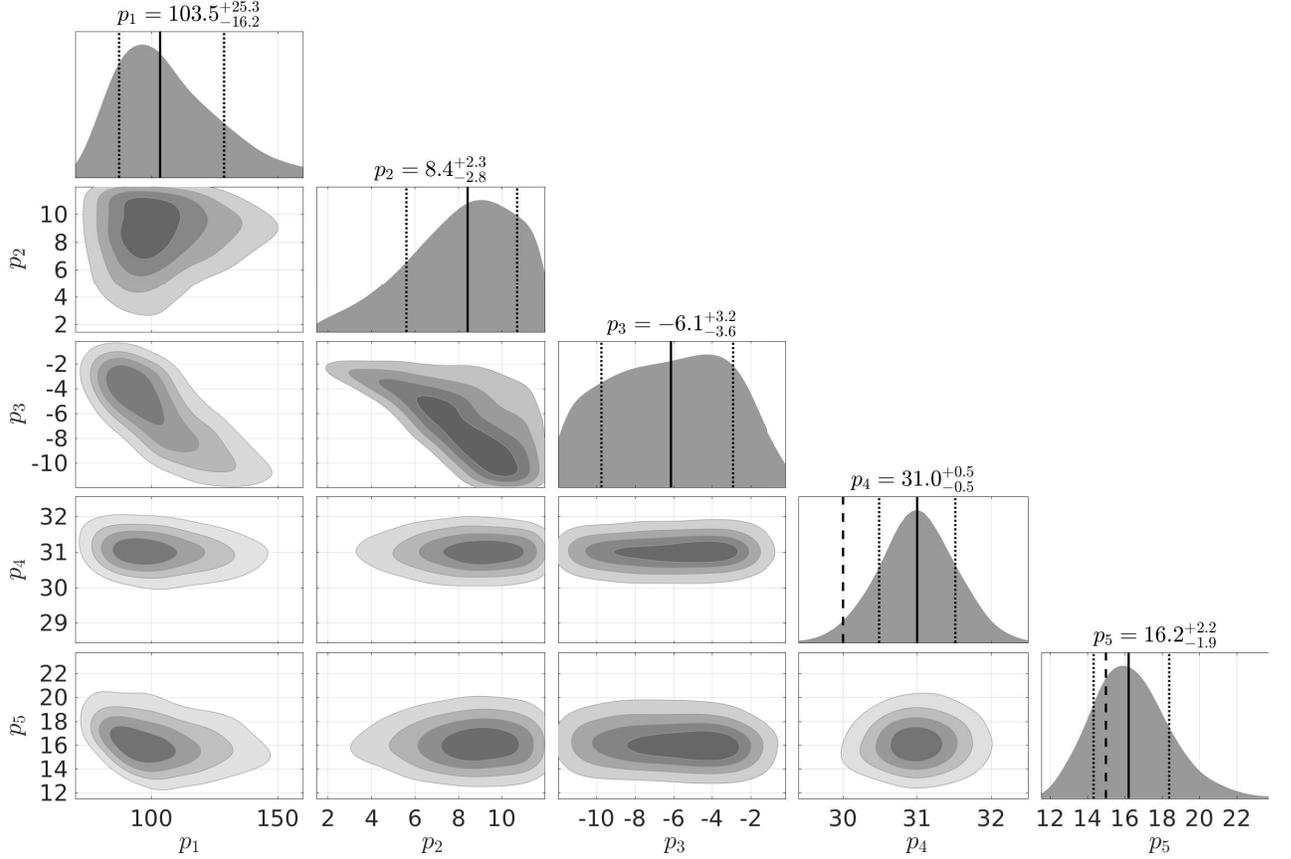}
  \caption{Corner plot showing the marginalized PDFs of the five parameters in the supergranulation example. The marginalized distribution for each parameter independently is shown in the histograms along the diagonal, and the marginalized 2D distributions as contour plots in the other panels. For each 1D histogram, the median of the PDF is the solid black line, and the dotted lines give the 68\,\% confidence interval, whose numerical values are provided at the top of each panel. The \rev{dashed black} lines for $p_4$ and $p_5$ are the known input parameter values for the two in common. The contour levels of the 2D joint probability densities are at 20\,\%, 40\,\%, 60\,\%, and 80\,\% confidence intervals.}
  \label{fig:corner_sg}
\end{figure}
 This article presents a comparison of six SOLA inversions (two flow components at three target depths) to the probabilistic one for a simulated supergranule model. For completeness, details of the inversions and how they are  compared are itemized here:

 \begin{itemize}
 \item The simulation domain is $100$\,Mm on a side horizontally, sampled evenly at $1/3$\,Mm. It extends to $-25$\,Mm at the bottom boundary. Cross correlations and travel times were measured in an area $50$\,Mm on a side at the same sampling. Each map is therefore $150\times 150$ points.

 \item Born sensitivity kernels are computed on the same horizontal grid as the travel times and they extend to 15\,Mm below the surface using 55 points in depth. The kernels use input model power spectra that have been Fourier filtered to separately retain the first 3 radial orders, including the $f$-mode. Everything is computed to match the details of the travel times as discussed in Section~\ref{sec:setup}.

 \item The SOLA inversion code is the faster formulation in Fourier space that can be run in parallel \citep{jackiewicz2012}. It includes a cross-talk parameter as in \citet{svanda2011}. Trade-off curves (L-curves) were studied to find the best regularization parameters. To obtain sensible results in terms of noise and misfit, the 3D Gaussian target functions have FWHM  on the order of 10\,Mm and 3\,Mm in the horizontal and vertical directions, respectively. In general, the SOLA inversion results are very similar to the ones published by \citet{dombroski2013} using an RLS inversion scheme.

 \item The probabilistic inversion has $2\times 10^5$ steps. Each of the five parameters was assigned 120 walkers. With thinning every five steps, the PDF was sampled $2\times 10^5/(120\times 5) = 333$ times per walker, as before.  More walkers were used in this inversion than the meridional example because the PDFs are more complicated, and it is suggested to use more walkers for good sampling. These values are sufficient to give good autocorrelation properties of the walker time series.

\item The data vector $\vec{d}$ is composed of only a small subset (12 of the 135) of the travel-time measurement maps. To speed up the computation, each of the  $150\times 150$ pixel maps was  binned down to only  $19\times 19$ pixels. The data vector is therefore 4332 measurements.  To match the size of the data vector, the forward vector resulting from the operation $g(\vec{m})$ uses kernels that are also binned down. This is noteworthy, in that the SOLA inversion uses the full maps. The probabilistic inversion is quite powerful with substantially less data. Of course, the 4332 measurements are not all completely independent, as described by the covariance matrix. 
   
\item The full noise covariance matrix used in the likelihood function (Equation~\ref{eq:like}) was constructed from smaller pairs of covariances ${\rm Cov}[\delta\tau_i,\delta\tau_j]$ and ordered to correctly match the travel-time data vector.

\item The uniform priors on the model are given in Table~\ref{tab:priors}. As is common \citep{foreman2013}, the walkers are initialized with values in a small Gaussian ball, centered somewhere within the prior bounds. The initial guess is not critical, as the walkers soon quickly explore the full parameter space after the burn-in stage.

\item The entire run on a two-core desktop machine (thus minimal parallelization) was about 80~minutes. The most expensive task is the computation of the forward travel times in Equation~\ref{eq:bornforward}, which takes about 0.2\,seconds each.

\end{itemize}

 \section{Supergranulation Flow Inversions}
 \label{app:flows}


A few more results are presented here to add to those in Section~\ref{sec:sgresults}. The corner plot for the probabilistic inversion is given in Figure~\ref{fig:corner_sg}.

\begin{figure}
  \centerline{
    \includegraphics[width=.5\textwidth]{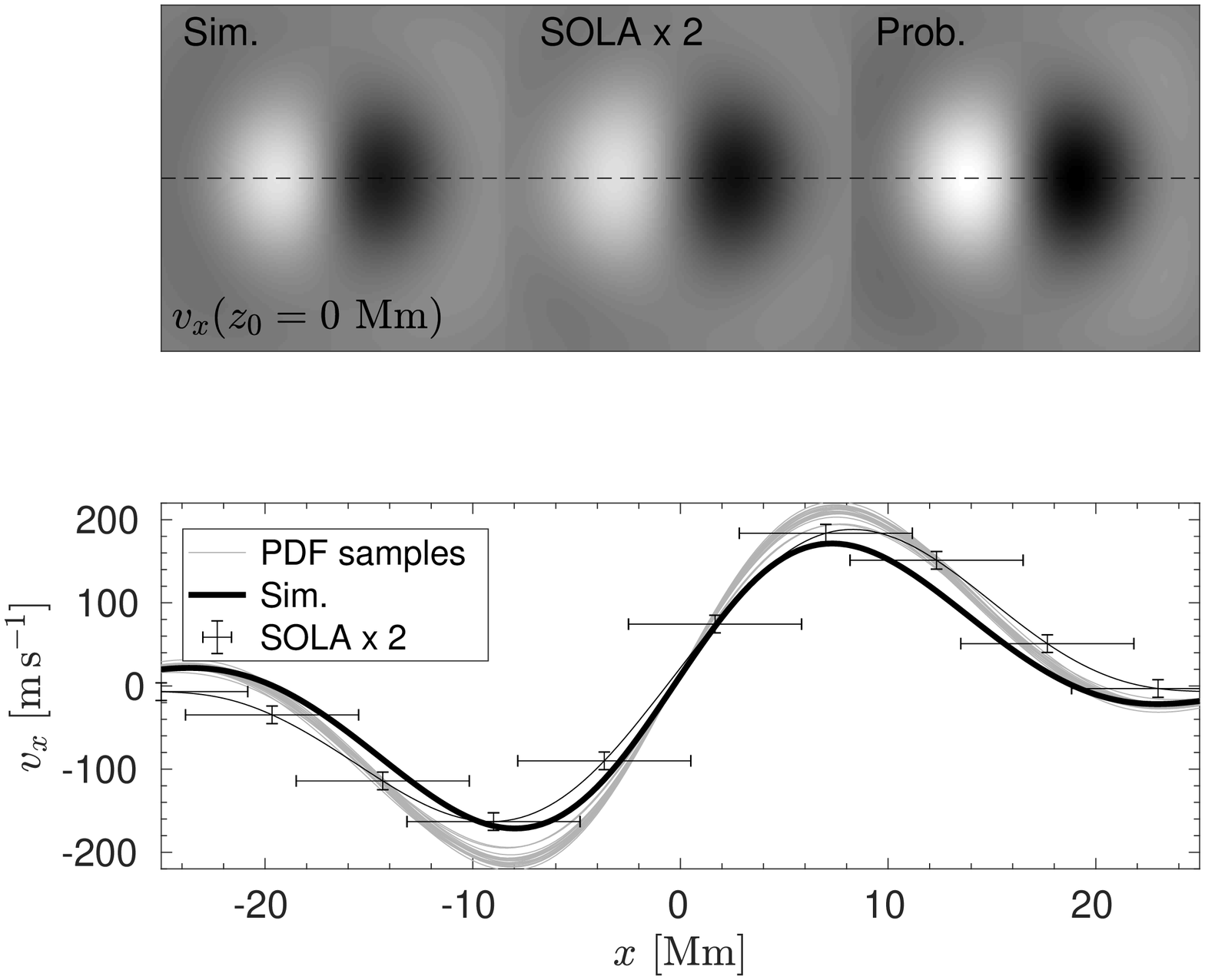}
    \includegraphics[width=.49\textwidth]{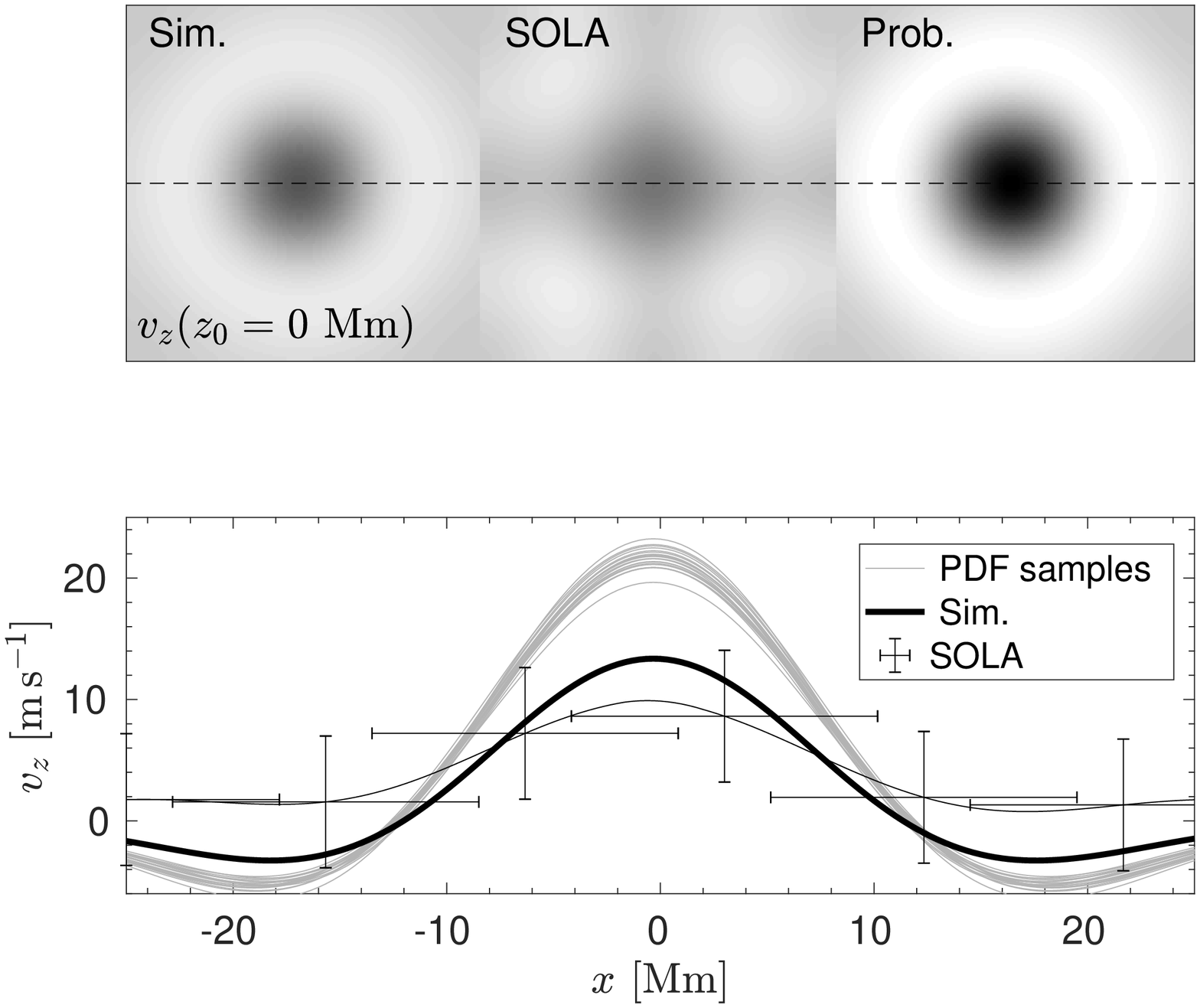}}
  \caption{Comparison of supergranulation inversion results at  the model photosphere. The left (right) panels are for the $v_x$ ($v_z$) inversion. The top rows are the flow fields for the simulation, the SOLA inversion, and the probabilistic inversion computed from the median of the PDF.  Darker shading corresponds to positive velocities, and the scale is the same in each set. The bottom panels are cuts through the supergranule at $y=0$, given by the dashed line in the top panels. Twenty random samples of the PDF are drawn and computed in the model space. A few representative points from the SOLA inversion are given with uncertainties.}
  \label{fig:v-0}
\end{figure}

The flow inversion comparison near the top of the simulation domain is shown in Figure~\ref{fig:v-0}. For the horizontal flows, the SOLA inversion underestimates the (smoothed) flow amplitude by about a factor of two, while the probabilistic inversion overestimates the amplitude by $15\,\%$. The profile structure is  primarily determined by parameters $p_4$ and $p_5$. The peak  vertical velocity is  underestimated by 35\,\% in the SOLA inversion, and it is overestimated by 60\,\% in the probabilistic inversion. The $f$-mode is generally the most sensitive at these layers, but its sensitivity to vertical flows is weak. This affects both inversions. $v_z^{\rm SOLA}$ also has artifacts in its flow structure.

\begin{figure}
  \centerline{
    \includegraphics[width=.5\textwidth]{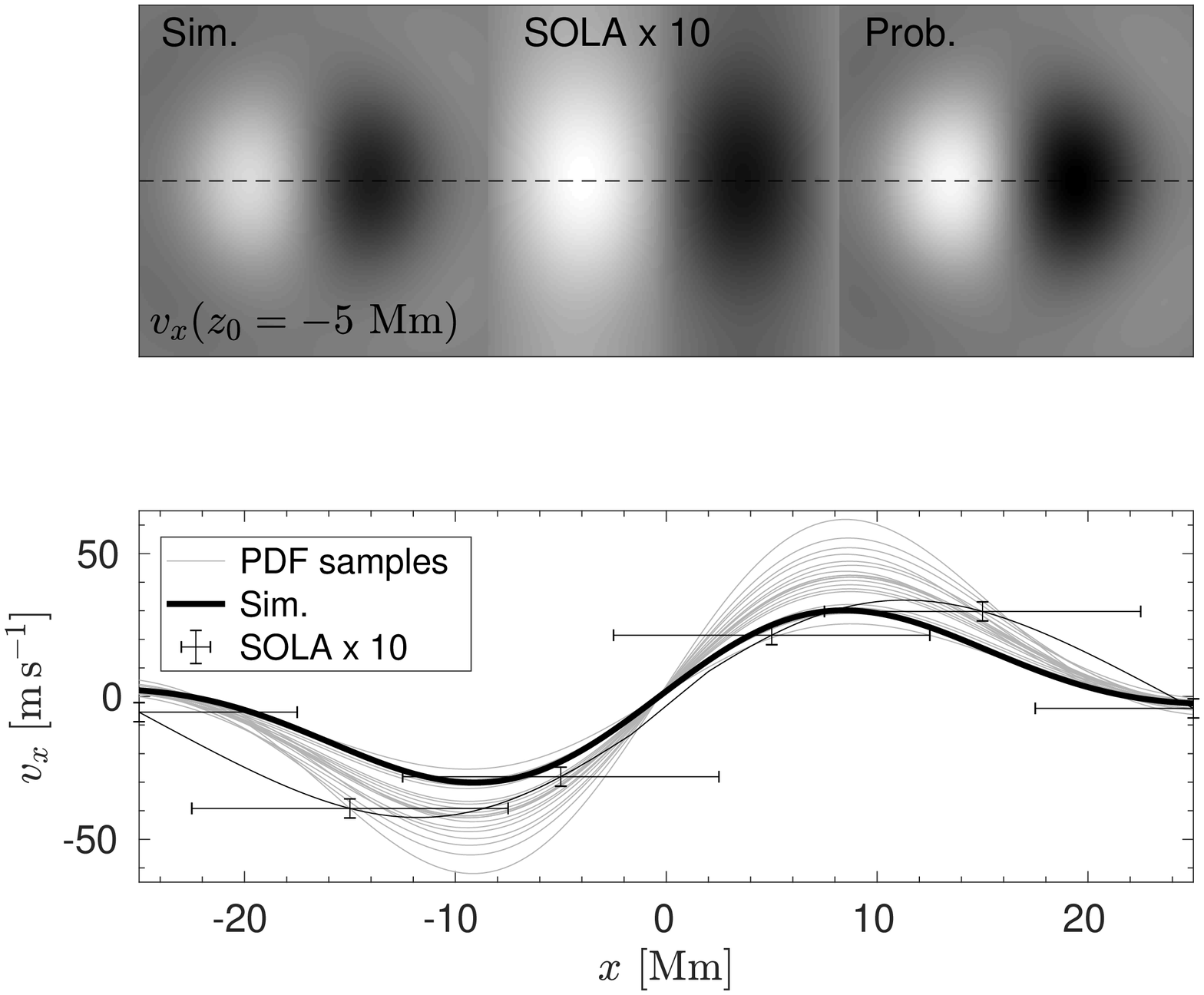}
    \includegraphics[width=.49\textwidth]{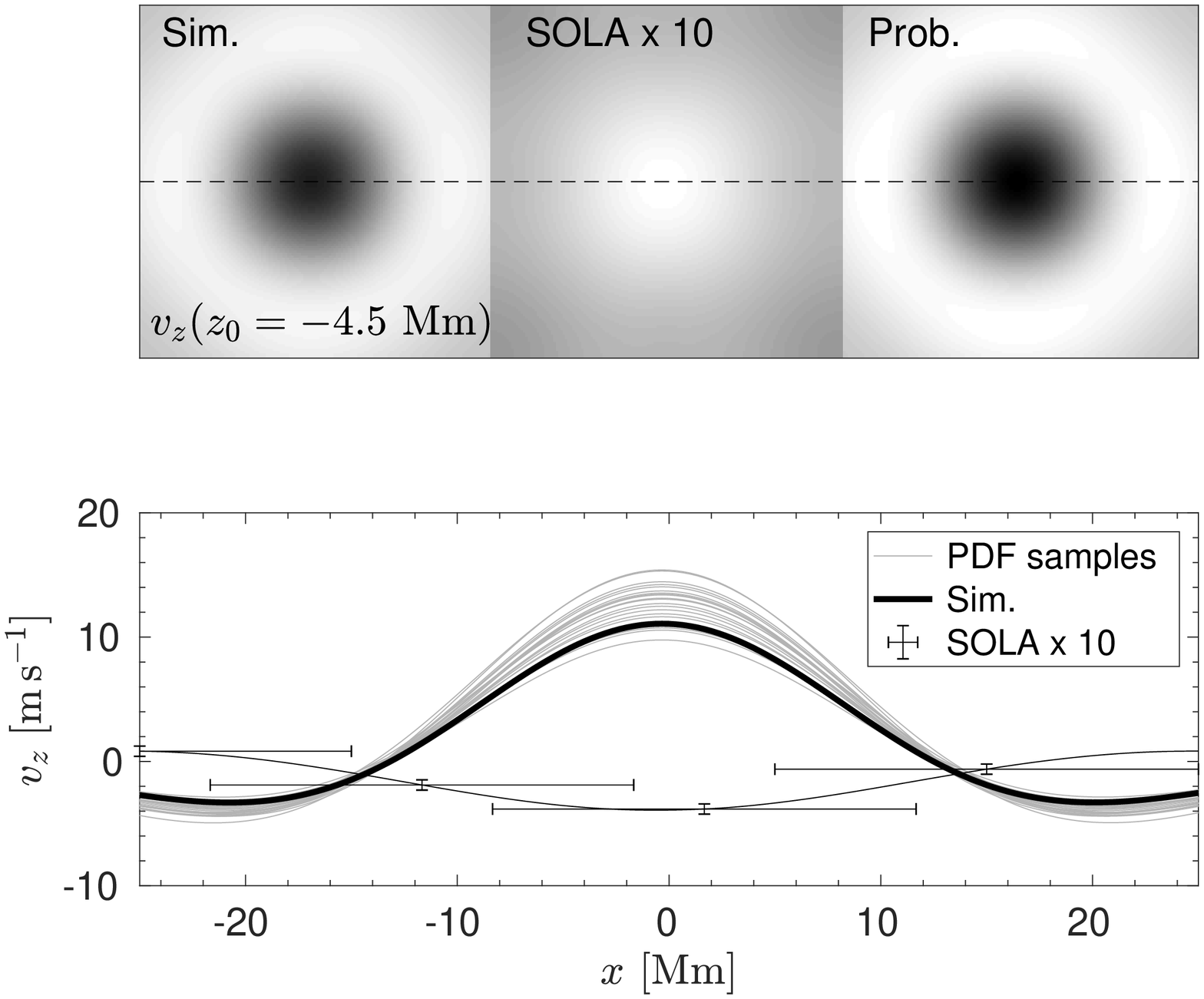}}
  \caption{Comparison of supergranulation inversion results near about 5\,Mm beneath the model photosphere. The left (right) panels are for the $v_x$ ($v_z$) inversion. The top rows are the flow fields for the simulation, the SOLA inversion, and the probabilistic inversion computed from the median of the PDF.  Darker shading corresponds to positive velocities, and the scale is the same in each set. The bottom panels are cuts through the supergranule at $y=0$, given by the dashed line in the top panels. Twenty random samples of the PDF are drawn and computed in the model space. A few representative points from the SOLA inversion are given with uncertainties.}
  \label{fig:v-5}
\end{figure}

A comparison near 5\,Mm beneath the surface is given in Figure~\ref{fig:v-5}. As is common in standard helioseismic inversions using typical data sets, the inferences at this depth are uninformative \citep{zhao2007}. The vertical velocity inferred is weak and is the wrong sign, which was also noticed in \citet{dombroski2013}. The probabilistic inversion provides a good estimate of the  flow structure at this depth.

\section{Why do probabilistic inversions work better?}

The comparisons of the different inversion methods reveal that the probabilistic ones are more robust at inferring the true flows
in the examples provided. To try to understand these results at a conceptual level, it is necessary to look at simpler examples that remove unneccesary complications.

We will focus on a problem that is basic, yet very similar to the form of helioseismic inversions -- recovering an image that has been blurred and corrupted with noise. Consider an initial image $x$ and the operation
\begin{equation}
  y = Ax + n,
\label{eq:form}
\end{equation}
where $A$ is some blurring operator and $n$ is independent Gaussian noise with zero mean [${\cal N}(0,\sigma^2)$]. $y$ is the resulting image that represents the ``data.'' This expression closely resembles the relationship between travel-time measurements and ($y$), sensitivity kernels ($A$),  internal flows ($x$), and the inherent noise in travel times ($n$).

\begin{figure}[t]
  \centering
  \includegraphics[width=.5\textwidth]{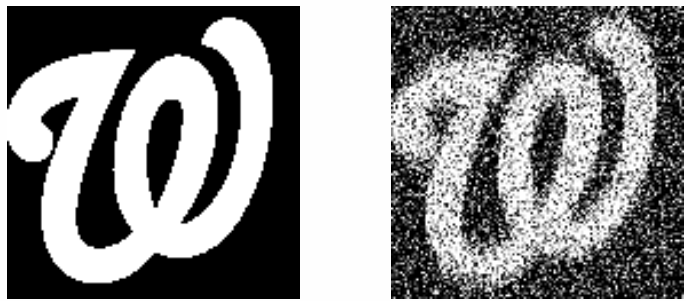}
  \caption{Left: A binary image ($x$) with $128^2$ pixels of $+1$ (white) and $-1$ (black). Right: Resulting image after blurring $x$ with a Gaussian kernel of standard deviation five pixels and adding Gaussian noise with zero mean and standard deviation $\sigma=1$. Note that the gray scale is truncated at $\pm\,1$, even though  $y$ takes on values larger than that.  The astute reader may recognize  a modified ``curly W'' logo of the Washington Nationals baseball team.}
  \label{fig:ls_examp}
\end{figure}

To make this even more straightforward, we will take  $x$ to be a binary image of a curly W letter, which  only has values of $+1$ (black) and $-1$ (white). An example of $x$ and the resulting data ($y$) with signal-to-noise ratio (SNR) of $1.5$ is shown in Figure~\ref{fig:ls_examp}.  The goal is to use the corrupted image on the right to try to infer the original image on the left. 

\subsection{SOLA-Type Solution}

One of the most straightforward ways of solving this image-recovery problem that most closely relates to the SOLA formulation described in the article is using a least-squares estimate. In fact, it is almost completely analogous to SOLA mathematically. The problem of estimating $x$ requires solving
\begin{equation}
  x_{\rm LS} = \arg\min_x \left\{ \left\Vert y - Ax\right\Vert^2 \right\}.
  \label{eq:ls}
\end{equation}
The least-squares solution can be found analytically as
\begin{equation}
  x_{\rm LS} = (A^{\rm T}\Sigma^{-1}A)^{-1}A^{\rm T}\Sigma^{-1}y,
  \label{eq:solls}
\end{equation}
where $\Sigma$ is the covariance matrix, which in this case is diagonal with value $\sigma^2$. Note the similarity to Equations~\ref{eq:cost} and \ref{mest}.

\begin{figure}[t]
  \centering
  \includegraphics[width=\textwidth]{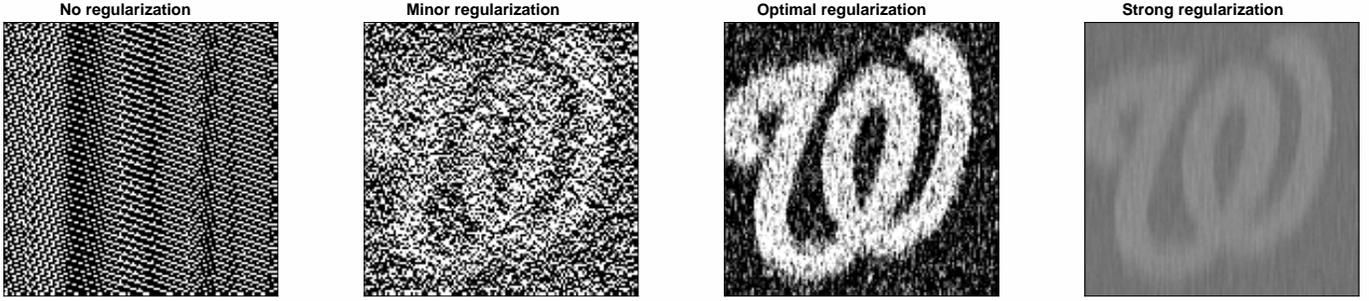}
  \caption{Least-squares solutions for the image reconstruction problem posed in Figure~\ref{fig:ls_examp} with $\sigma=1$. Each panel represents the solution using various amounts of regularization, from none (left) to overly regularized (right). The values are clipped to $\pm\,1$.}
  \label{fig:ls_array}
\end{figure}

The solution computed using Equation~\ref{eq:solls} is shown in Figure~\ref{fig:ls_array} in the left panel. Evidently, the solution is very unstable, as the matrix A is ill-conditioned and sparse with small values, which causes the inverse (if it exists, as it does in this case) to blow up. There are methods to avoid this behavior by \textit{regularizing} the problem. Instead of solving Equation~\ref{eq:ls}, one rewrites the problem as
\begin{equation}
  x_{\rm LS} = \arg\min_x \left\{ \left\Vert y - Ax\right\Vert^2 + \alpha\left\Vert x\right\Vert^2\right\},
  \label{eq:rls}
\end{equation}
where solutions can be controlled by a regularization parameter $\alpha$. The analytic solution in this case is
\begin{equation}
  x_{\rm LS} = (A^{\rm T}A+\alpha I)^{-1}A^{\rm T}y,
  \label{eq:solrls}
\end{equation}
where $I$ is the identity matrix, and the covariance matrix has been omitted for clarity.  The regularizing parameter essentially controls the smoothness of the reconstructed image. When small, the solution fits the data well at the expense of amplified noise and not being smooth; when large, the solution is smooth but does not fit the data well. The key is that the solution either resembles $x$ in some fashion but is very noisy, or is smoother with a much smaller amplitude.

This is demonstrated in Figure~\ref{fig:ls_array} in the final three panels, which are computed using Equation~\ref{eq:solrls}. When there is minor regularization, the image shows the curly W, but with rather high-amplitude pixels caused by the noise amplification. The maximum values reach about 20, and the rms over all pixels is 4 (the rms of the original image is 1). Even though this solution is  technically the best fit to the data, it is very noisy. As is common in helioseismic inversions, an ``L''-curve can be computed from the series of  solutions based on the regularization trade off. The optimal regularization solution in the third panel is where the bend in the ``L''-curve occurs in this example. It represents a compromise between smoothness and noise. The final panel is overly regularized and smooth, with the letter almost completely washed out. Its values are between $\pm\,0.15$.

In this example, as in SOLA inversions for  helioseismology, a set of solutions is calculated based on different levels of regularization. The ``best'' solution is not chosen arbitrarily, but it is chosen with little guidance about the physics of the problem. It is the solution that ``looks good,'' and is mostly due to the addition of an unphysical regularization parameter. If none of the solutions look good, there are not many options left.


\subsection{Probabilistic Solution}

The image reconstruction inverse problem can also be solved in a probabilistic fashion using the Bayesian formalism. One  strategy is to try to determine, for every pixel in the noisy $y$ image, if it came from a $+1$ in $x$ or from a $-1$ in the original.

First, we need a likelihood function and priors. The likelihood function of the data ($y$) given some model image estimate ($m$)  of the true image  is
\begin{equation}
  L(y|m,\sigma) \propto\exp\left[ -\frac{1}{2\sigma^2}\sum_{ij}(y_{ij} - m_{ij})^2\right],
\end{equation}
where the sum is over all $M$\,x\,$N$ pixels in the image. The value of any $(i,j)$th pixel in our model can only take on $+1$ or $-1$.

The Bayesian framework allows one to use various priors. A uniform prior means that any model image is equally probable, and each pixel is independent from every other pixel. In this case, uniform priors do not work so well. The reason is that for noisy images made up with pixels of large amplitude, the probability that any pixel was derived from  $+1$ is very similar to the probability it came from $-1$. From testing, this does not help significantly reduce the noise on the reconstructed image.

However, using \textit{a priori} knowledge that the goal is to reconstruct the binary image of a letter, where pixels of the same value tend to be clumped together, we can compute a more useful prior in the following fashion.  A pixel is compared to its four neighbors. If this pixel is embedded near pixels of the same sign, the probability that it has that sign  is larger than otherwise. Therefore, the prior for any given pixel can be expressed as
\begin{equation}
  \rho(m_{ij})\propto\exp\left( - J m_{ij}^*\right),
\end{equation}
where $m_{ij}^*$ is simply the number of neighboring pixels around $m_{ij}$ that have a different sign. $J$ is a (positive) tunable coupling constant. A large $J$, for example, produces big blobs of white or black pixels. In these examples we use $J=0.5$, which tends to give reasonable results. This prior is motivated by the Ising model used in the study of magnetism in solid-state physics \citep{brush1967} and now utilized in many areas of MCMC sampling \citep[\textit{e.g.}][]{donner2017}. The Ising prior would be, to our knowledge, challenging or impossible to implement in the least-squares inversion framework.

The posterior probability function can now be computed and sampled using MCMC.  A few details of how the Markov chains proceed are worthwhile to explain.  The  model trial image is initialized to all ones. The algorithm iterates pixel-by-pixel. For each pixel, the sign gets flipped as a proposal, and the probability of accepting this new state of the pixel is computed as the ratio of the PDF of the new state to the PDF of the original state.  If the ratio is greater than 1, the new state is accepted. If not, a random number is drawn between 0 and 1, and if the number is less the ratio, the new state is accepted anyway. This is a familiar MCMC update procedure (\textit{e.g.} Metropolis-Hastings), even simpler than the one used in the rest of this article.

\begin{figure}[t]
  \centering
  \includegraphics[width=.98\textwidth]{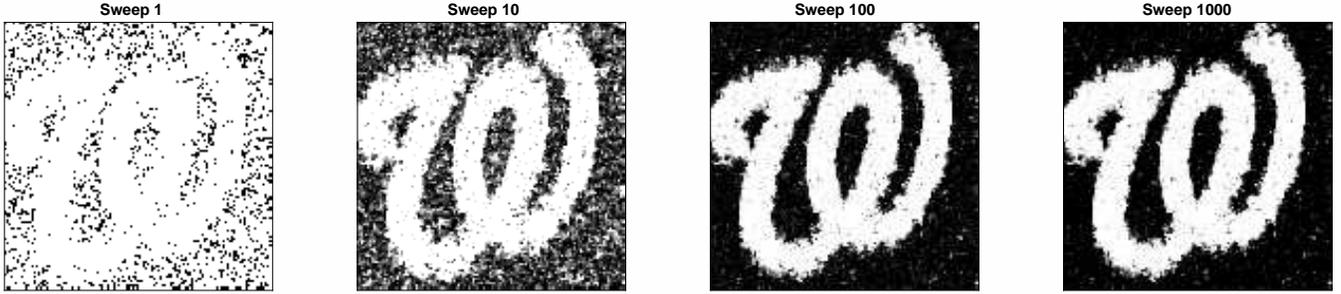}
  \caption{Probabilistic solution for the image reconstruction problem posed in Figure~\ref{fig:ls_examp}. Each panel represents the mean solution after the number of sweeps indicated, where each sweep is 10,000 realizations. The reconstructed images can take on any values in the range $\pm\,1$.}
  \label{fig:mc_array}
\end{figure}

Figure~\ref{fig:mc_array} shows a few stages, or sweeps, of the Bayesian inversion as it samples the PDF. A snapshot of the current MCMC state is taken every 10,000 iterations, and the $n$th sweep is an average over $n$ snapshots. It is analagous to studying means or other modes of the PDF, rather than the value of a parameter at any state (e.g., the value of any walker at some arbitrary iteration is not very informative).   During the first sweep, the model image is initialized to all ones, and after 10,000 proposed pixel flips, the curly W is only marginally visible. Each pixel may only have been visited on average once, or less ($128^2=16,384$). At sweep 10 the algorithm has gone through $10^5$ iterations, and the reconstruction is getting smoother.  After $10^7$ iterations, the final recovered image is quite similar to the original one.

\subsection{Summary}

The Bayesian framework is more flexible and works better than least-squares algorithms for problems of the form of Equation~\ref{eq:form}, whose  form is the generally the same as in helioseismology. Not only do the probablistic inversions work better, they allow for potentially many \textit{different} types of solutions since they explore the full posterior probability, whose landscape may be complex. The least-square solution only gives one class of solutions, which are often very sensitive to numerical instabilities and noise.  With different prior assumptions, other sets of solutions can be explored with the Bayesian method. 

\subsection{Animations}
A few animations of the inversions as they proceed can be found at \url{http://astronomy.nmsu.edu/jasonj/MCMC/}. The three animations show example reconstructions using both methods for different noise levels ${\cal N}(0,\sigma^2)$; for $\sigma^2=0$, $\sigma^2=2.0$, and $\sigma^2=5$. The last case only has an SNR$=0.3$. The animation updates after each sweep of 100,000 iterations. The top panels are the original image and the blurred and noisy image. The bottom left is the current MCMC state, which is a binary image. The bottom center is the image at the $n$th sweep. The bottom right shows the least-squares solution for the given regularization (smoothing), whos value increases with time. All panels have a grayscale with limits set to $\pm 1$.

The example with no added noise is interesting. The image is blurred only. The least-squares solution can recover the blurred image, but not the source image, as the Bayesian inversion does. 

%
%

\end{document}